\documentclass[prd,twocolumn,showpacs,nofootinbib,amssymb]{revtex4}

\usepackage{graphicx}% Include figure files
\usepackage{epsfig}
\usepackage{dcolumn}% Align table columns on decimal point
\usepackage{bm}% bold math
\usepackage{color}
\usepackage{subfigure}
\newcommand{\comment}[1]{}
\newcommand{\newc}{\newcommand}
\def\Journal#1#2#3#4{{#1} {\bf #2}, #3 (#4)}
\def\emvide{{\em vide}~}
\def\NPB{{ Nucl. Phys.} B}
\def\PLB{{ Phys. Lett.}  B}

\def\PRD{{ Phys. Rev.} D}
\def\ZPC{{ Z. Phys.} C}

\def\JHEP{ JHEP}

\def\thetab0{\theta_{B_0}}

\def\lsim{\ ^<\llap{$_\sim$}\ }

\def\r2{\sqrt 2}
\def\beq{\begin{equation}}
\def\eeq{\end{equation}}
\def\beqn{\begin{eqnarray}}
\def\eeqn{\end{eqnarray}}
\def\sinW2{\sin^2\theta_W}

\def\mz2{M_{z}^2}
\def\c2b{\cos 2\beta}

\def\mz{M_Z}

\def\sec2w{sec^2\theta_W}
\def\gmin2{(g-2)_\mu}

\def\dh {\partial }
\newc{\wt}{\widetilde}
\newc{\ra}{\rightarrow}
\newc{\s}{\smallskip}
\newc{\nn}{\noindent}
\newc{\non}{\nonumber}
\def \chonep{{\wt\chi_1}^{+}}
\def \chonem{{\wt\chi_1^-}}
\def \chonep2{{\wt\chi_2^+}}
\def \chonem2{{\wt\chi_2^-}}

\def \wtchi  {{\wt \chi }}
\begin{document}
\begin{flushright}
\hspace*{6in}
\mbox{LPT-Orsay-10-22}
\end{flushright}

%\preprint{LPT-Orsay-10-22}

\title{One-loop contribution to the neutrino mass matrix in NMSSM  
with right-handed neutrinos and tri-bimaximal mixing}

\author{Debottam Das}%
\email{debottam.das@th.u-psud.fr}
\affiliation{Laboratoire de Physique Th\'eorique, UMR 8627,
Universit\'e de Paris-Sud 11, B\^atiment 210, \\ 91405 Orsay Cedex,
France}
\author{Sourov Roy}%
\email{tpsr@iacs.res.in}
\affiliation{Department of Theoretical Physics and Centre for Theoretical
Sciences, Indian Association for the Cultivation of Science, 2A $\&$ 2B Raja
S.C. Mullick Road, Kolkata 700 032, India}
\date{\today}
\begin{abstract}
Neutrino mass patterns and mixing have been studied in the context of 
next-to-minimal supersymmetric standard model (NMSSM) with three gauge singlet 
neutrino superfields. We consider the case with the assumption of R-parity 
conservation. The vacuum expectation value of the singlet scalar field $S$ of NMSSM
induces the Majorana masses for the right-handed neutrinos as well as the usual $\mu$-term.
The contributions to the light neutrino mass matrix at the tree level as well as one-loop level 
are considered, consistent with the tri-bimaximal pattern of neutrino mixing. Light neutrino 
masses arise at the tree level through a TeV scale seesaw mechanism involving the 
right-handed neutrinos. Although all the three light neutrinos acquire non-zero masses at 
the tree-level, we show that the one-loop contributions can be comparable in size under certain 
conditions. Possible signatures to probe this model at the LHC and its distinguishing features compared
to other models of neutrino mass generation are briefly discussed.
\end{abstract}


%\keywords{Suggested keywords}

\maketitle

\section{Introduction}
Several mechanisms of the generation of neutrino masses and mixing in the
context of a supersymmetric model have been explored in various works. One of
the most popular attempts in this direction is to relax the assumption
of R-parity conservation in the minimal supersymmetric standard model (MSSM) 
by including explicit bilinear and/or trilinear R-parity violating interactions in the superpotential 
and the scalar potential\cite{r-parity-early-references,r-parity-review}. One can also consider models with
spontaneous R-parity violation \cite{romao-santos-valle, giudice-masiero-pietroni-riotto, umemura-yamamoto} 
via a singlet sneutrino vacuum expectation value. The low energy limit of such models, where the singlet sneutrino 
field is decoupled, can be thought of as the bilinear R-parity violating scenario. Thus there are several possibilities 
within the context of R-parity violation in MSSM. In fact, each of them has been studied in detail in connection 
with the observed neutrino mass patterns and mixing as provided by the neutrino oscillation experiments. 
The possible collider signatures of R-parity violating models have also been studied in great details and correlation 
between neutrino mixing angles and the decay branching ratios of the lightest supersymmetric particle (LSP) have been 
obtained \cite{gonzalez-garcia-romao-valle, adhikari-mukhopadhyaya, hirsch-vicente-porod, Roy-Mukhopadhyaya-Vissani, 
choi-chun-kang-lee, romao-diaz-hirsch-porod-valle, datta-mukhopadhyaya-vissani, porod-hirsch-romao-valle, chun-jung-kang-park, 
jung-kang-park-chun}. 

Another interesting and well studied procedure of small neutrino mass 
generation in a supersymmetric model, with the observed mixing pattern, is the 
seesaw mechanism\cite{original-seesaw,other-early-seesaw} with the introduction of right-handed neutrino 
superfields\cite{grossman-haber,susy-seesaw,davidson-king}. In order to generate small neutrino masses, 
one introduces $\Delta L$ = 2 heavy Majorana mass terms in the superpotential in addition to 
the trilinear lepton-number conserving  Yukawa interactions involving the 
right-handed neutrino superfields. As long as the neutrino Yukawa couplings are
of order one, light neutrino masses $\sim 10^{-2}$ eV require the Majorana
masses to be $\sim 10^{15}$ GeV or so. However, such a high seesaw scale 
is difficult to probe at the LHC or future linear collider experiments. A viable 
alternative is to look at {\it TeV-scale} seesaw mechanism where small active neutrino 
masses are generated with the help of neutrino Yukawa couplings as small as $10^{-6}$ 
(same as the electron Yukawa coupling) and this makes the Majorana mass scale of the 
right-handed neutrino of the order of $\sim$ TeV plausible. This gives one an opportunity
to test the seesaw models at the LHC. The signatures of TeV scale supersymmetric seesaw models  
will be briefly outlined later along with a discussion of the signatures of R-parity-violating models. 

On the other hand, MSSM is plagued by the so-called ``$\mu$-problem" which asks 
the question that why the scale of the supersymmetry preserving $\mu$-term
should be of the same order as the soft supersymmetry breaking terms, which are
of the order of TeV. One of the possible solutions to this problem is the next-to-minimal 
supersymmetric standard model (NMSSM), where a standard model singlet 
superfield ($ \hat S$) is introduced to the MSSM superfields with a coupling
$\lambda { \hat S} { \hat H}_u {  \hat H}_d$ in the superpotential
(for review and phenomenology see\cite{NMSSM_review1,NMSSM_review2}). The scalar 
component of ${ \hat S}$ gets, in general, a non-zero vacuum expectation value (VEV)
of the order of $\sim$ TeV, as long as the soft mass parameters
corresponding to the singlet scalar field are in the same range. 
This solves the ``$\mu$-problem" because the $\mu$-parameter generated in 
this way has the right order of magnitude if one considers a coupling $\lambda 
\sim {\cal O}(1)$. In order to generate active neutrino masses and appropriate 
mixing in the neutrino sector one either includes R-parity violation in the 
superpotential\cite{chemtob-pandita,abada-bhattacharyya-moreau} and the scalar 
potential or introduces gauge-singlet neutrino superfields $\hat N_i$ with 
appropriate couplings with the MSSM superfields and the singlet superfield 
${ \hat S} $\cite{Kitano-2001}. In the latter case, the gauge-singlet neutrino superfields 
$\hat N_i$ can have Majorana masses around the TeV scale if there is a coupling of the 
type $\kappa  \hat N_i^2  \hat S$ in the superpotential. When the scalar component of 
${ \hat S}$ gets a VEV of the order of TeV scale, the right handed neutrinos also acquire 
an effective Majorana mass around the TeV values as long as the dimensionless coupling 
$\kappa$ is order one\cite{Kitano-2001}. Here it is assumed 
that the superpotential has a discrete $Z_3$ symmetry which forbids the 
appearance of bilinear terms in the superpotential \cite{ellis89}.

In this study, within the framework of this TeV scale seesaw model mentioned above, 
we calculate the one-loop contributions to the neutrino 
mass matrix with R-parity conservation and study the effect of these 
contributions to the neutrino mass patterns and mixing angles. In other 
words, we consider the case where only the scalar field corresponding to the singlet 
superfield $\hat S$ gets a non-zero VEV along with the neutral Higgs fields. 
We will show later that these one-loop contributions 
can be significant and can change the region of parameter space allowed by the three-flavor global 
neutrino data in comparison to the tree level results. 

The plan of the paper is as follows. In Sec.II we will provide a discussion on the three-flavor neutrino 
mixing and illustrate the general pattern of our analysis that we are going to follow. Sec.III describes 
the model along with the minimization conditions of the neutral scalar potential. One-loop contributions 
to the neutrino mass matrix in the R-parity conserving scenario and the resulting neutrino mass patterns, 
which satisfy the three flavor global neutrino  data, are discussed in Sec.IV with numerical results. In 
Sec.V we outline the possible ways to probe this model at the LHC and present a short critical discussion 
of the signatures of neutrino mass models involving spontaneous and/or bilinear R-parity violation. 
We summarize in Sec.VI with possible future directions.

\section{Neutrino mixing}
\label{neumixing}
The solar, atmospheric, accelerator, and reactor neutrino
experiments have shown strong evidence in favor of non-zero neutrino masses
and mixing angles\cite{Strumia}. In addition, there is an upper bound 
on the sum of neutrino mass eigenvalues $\sim$1 eV from cosmological 
observations\cite{neutrino_sum}. The bound on the 11-element of the neutrino 
mass matrix resulting from the non-observation of neutrinoless double beta 
decay is $\le$0.3 eV\cite{neutrino_mass}. The global 3-flavor fits of various 
neutrino oscillation experiments point toward the following 3$\sigma$ ranges of 
the neutrino oscillation parameters, namely the two mass-squared differences 
and three mixing angles%\cite{neu-recent-data, PDG, data}: 
\cite{new-neu-recent-data}:
\begin{eqnarray} 
&&\sin^2\theta_{12} = 0.25-0.37 ,\ \sin^2\theta_{23} =0.36-0.67 ,\nonumber \\
&&\sin^2\theta_{13} \le 0.056 \nonumber \\
&&\Delta m_{21}^2 =(7.05-8.34)\times 10^{-5} \; {\rm eV}^2\;,\nonumber \\ 
&&|\Delta m_{31}^2| =(2.07-2.75)\times 10^{-3}\; {\rm eV}^2\;,
\label{eq:data}
\end{eqnarray}
where $\Delta m^2_{ij} \equiv m^2_i - m^2_j$.
One can see from these numbers that there are two large mixing angles and 
one small mixing angle among the three light neutrinos with a mild hierarchy 
between the mass eigenvalues. 

The three flavor neutrino mixing matrix $U$ can be parametrized as follows, 
provided that the charged lepton mass matrix is already in the diagonal form and
the Dirac as well as Majorana phases are neglected:

%\begin{widetext}
\begin{eqnarray}
U = \left( \begin{array}{ccc}
        c_{12}c_{13} & s_{12}c_{13} & s_{13} \\
        -s_{12}c_{23}-c_{12}s_{23}s_{13} &
        c_{12}c_{23}-s_{12}s_{23}s_{13} & s_{23}c_{13} \\
        s_{12}s_{23}-c_{12}c_{23}s_{13} &
        -c_{12}s_{23}-s_{12}c_{23}s_{13} & c_{23}c_{13}
     \end{array} \right),\nonumber \hskip -0.8 cm \\ 
\end{eqnarray}  
%\end{widetext}
where $c_{ij} = \cos\theta_{ij}$, $s_{ij} = \sin\theta_{ij}$ and $i,j$ run 
from 1 to 3.

The mixing angle data coming from solar, atmospheric and reactor sector 
indicate that $\theta_{12} \approx 34^\circ$, $\theta_{23} \approx 45^\circ$, 
and $\theta_{13} \le 13^\circ$. This is popularly known as the bilarge pattern
of neutrino mixing. In order to understand the consequences of such mixing in 
the {\it zeroth} order, one considers the tri-bimaximal structure of the 
neutrino mixing\cite{hps} where $\theta_{23} = \frac{\pi}{4}$, $\theta_{13} 
= 0$ and $\sin\theta_{12} = \frac{1}{\sqrt{3}}$. 

With this tri-bimaximal pattern, the unitary neutrino mixing matrix turns out 
to be 
\begin{equation}
U_\nu = \left( \begin{array}{ccc}
        \sqrt{\frac{2}{3}} & \frac{1}{\sqrt{3}} & 0 \\
        -\frac{1}{\sqrt{6}} & \frac{1}{\sqrt{3}} & \frac{1}{\sqrt{2}} \\
        \frac{1}{\sqrt{6}} & -\frac{1}{\sqrt{3}} & \frac{1}{\sqrt{2}}
        \end{array} \right).
\end{equation}

Considering $m_1$, $m_2$ and $m_3$ as the three light neutrino mass 
eigenvalues, we use the matrix $U_\nu$ to obtain the neutrino Majorana mass 
matrix in the flavor basis as 
\begin{widetext}
\begin{eqnarray}
m_\nu &=& U_\nu \left( \begin{array}{ccc}
        m_1 & & \\
         & m_2 & \\
         & & m_3
        \end{array} \right) U^T_\nu
                \nonumber \\
      &=& \left( \begin{array}{ccc}
        \frac{1}{3}(2m_1+m_2) & \frac{1}{3}(-m_1+m_2) &
        \frac{1}{3}(m_1-m_2) \\
        \frac{1}{3}(-m_1+m_2) & \frac{1}{6}(m_1+2m_2+3m_3) &
        \frac{1}{6}(-m_1-2m_2+3m_3) \\
        \frac{1}{3}(m_1-m_2) & \frac{1}{6}(-m_1-2m_2+3m_3) &
        \frac{1}{6}(m_1+2m_2+3m_3)
        \end{array} \right).
\label{neutribi}
\end{eqnarray}
\end{widetext}
We can see that a particular structure of neutrino mass matrix emerges from
the requirement of tri-bimaximal mixing, in terms of the neutrino mass 
eigenvalues. Given a specific model for generating the neutrino mass matrix,
one can easily connect the model parameters with the neutrino mass eigenvalues
with the help of Eq.(\ref{neutribi}). This way one can study the normal, 
inverted or quasi-degenerate mass pattern of the light neutrino mass eigenvalues and 
try to see the requirement on the model parameters to produce the tri-bimaximal
pattern of neutrino mixing. In this work, we will try to explore the 
next-to-minimal supersymmetric standard model (NMSSM) where neutrino mass is 
generated because of the introduction of three right-handed neutrino 
superfields with the possible interaction terms. Though the assumption of tri-bimaximal
mixing in the neutrino sector is not generic, in the present context it is quite 
illustrative in studying the role of the soft SUSY breaking parameters on the neutrino mass 
eigenvalues. At the same time, the acceptable domain of the soft parameters
consistent with neutrino mass eigenvalues and tri-bimaximal mixing angles 
would hardly change with any small shift in $\theta_{13}$. 

As mentioned in the introduction, this model was proposed in Ref.\cite{Kitano-2001} where 
the case with spontaneous violation of R-parity was studied with possible implications 
on neutrino mass eigenvalues and mixing angles at the tree level. In the present 
study we shall consider the case when R-parity is conserved and the neutrino mass 
generation at the tree level is entirely due to the seesaw mechanism involving the 
TeV scale right handed neutrinos. Our aim would be to see if this model can produce 
the acceptable neutrino mass eigenvalues and mixing angles when the neutrino mass matrix 
receives contributions at the tree as well as one-loop level. An attractive 
feature of this model is that, the right handed sneutrino in the form of LSP 
may become a valid cold dark matter candidate of the universe\cite{cerdenoall}.

This model can also 
accommodate spontaneous CP and R-parity violation simultaneously. In that case, the 
neutrino sector is CP violating and the resulting effects on the neutrino masses and 
mixing angles  were studied in Ref.\cite{katri-mariana-timo}. 
Similarly, spontaneous R-parity violation motivated by a flavor symmetry 
may produce tri-bimaximal mixing pattern in the neutrino sector\cite{manimala}.
However, in the present context we consider the case where 
neutrino sector conserves CP symmetry along with R-parity.   

There have been some other studies which address the neutrino 
experimental data in some other extensions of NMSSM. One of these proposals is discussed in 
Ref.\cite{chemtob-pandita}, where the effective bilinear R-parity breaking 
terms are generated through the vacuum expectation value of the scalar 
component of the singlet superfield ${ \hat S}$. In this case, only one neutrino
mass is generated at the tree level whereas the other two masses are generated
at the one-loop level. In another model\cite{abada-bhattacharyya-moreau}, non-zero 
masses for two neutrinos are generated at the tree level by including explicit
bilinear R-parity violating terms along with the R-parity breaking term 
involving ${ \hat S}$. It is interesting to note that, the R-parity violating NMSSM model 
may offer a valid dark matter candidate in the form of gravitino as the R-parity violating 
decay channels of the gravitino are extremely suppressed because of weak gravitational 
strength\cite{moreau-dark}.

In another class of models, gauge-singlet neutrino superfields were introduced
to solve the $\mu$-problem, which can simultaneously address the desired 
pattern of neutrino masses and mixing\cite{munoz}. The detailed study of
neutrino masses and mixing in this model was presented in Ref.\cite{ghosh-roy} 
and the correlations of the lightest neutralino decays with neutrino mixing angles were 
discussed. Subsequently the dominant one-loop contributions towards the tree level neutrino 
masses have also been presented\cite{ghosh-roy-loop}. Similar analyses for one and two 
generations of gauge-singlet neutrinos were presented in Ref.\cite{hirsch-vicente} and some 
other phenomenological implications, in particular the possible signatures 
at LHC were addressed. Neutrino masses consistent with different hierarchical 
scenarios and tri-bimaximal neutrino mixing can also be generated in an R-parity violating 
supersymmetric theory with TeV scale gauge singlet neutrino superfields, where the $\mu$-term 
was not generated by the vacuum expectation values of the singlet sneutrino 
fields \cite{Mukhopadhyaya:2006is}. Another interesting avenue in this direction is to study the 
role of possible higher dimensional supersymmetry breaking operators in the hidden sector which 
may render the TeV scale soft SUSY breaking trilinear and bilinear couplings involving the sneutrinos 
to produce the observable mass and mixing angles for the neutrinos\cite{higher-dimension}.
\section{The model and minimization conditions}
In this section we review the model along the lines of 
Ref.\cite{Kitano-2001} and discuss its important characteristics. 
We introduce the singlet superfield $ \hat S$ along with three right-handed 
neutrino superfields $ \hat N_i$. The superfields $ \hat N_i$ are odd and the superfield 
$ \hat S$ is even under R-parity. The most general superpotential consistent with 
R-parity conservation is
\begin{eqnarray}
W&=&W_{\rm NMSSM}+W_{\it Singlet}\ , 
\label{Supot}
\end{eqnarray}
where
\begin{eqnarray}
W_{\rm NMSSM}&=&f^d_i( \hat H_d  \hat Q_i)  \hat D_i 
              +f^u_{ij} ( \hat Q_i  \hat H_u)  \hat U_j \nonumber \\
              &+& f^e_i ( \hat H_d \hat L_i)  \hat E_i +\lambda_H ( \hat H_d  \hat H_u) \hat S  
%\ , \nonumber \\
	      +\frac{\lambda_s}{3!}  \hat S^3 \ , \\	
W_{\it Singlet}&=&f^\nu_{ij}( \hat L_i  \hat H_u)  \hat N_j
         +\frac{{\lambda_N}_i}{2} \hat  N_i^2  \hat S. 
         \label{model-exp}
\end{eqnarray}
Here  $ \hat H_d$ and $ \hat H_u$ are down-type and up-type Higgs superfields, respectively.
The $ \hat Q_i$ are doublet quark superfields, $ \hat U_j [ \hat D_j]$ are singlet up-type 
[down-type] quark superfields. The $\hat L_i$ are the doublet lepton superfields, 
and the $ \hat E_j$ are the singlet charged lepton superfields. The indices $i,j = 
1,2,3$ are generation indices. Note that we have imposed a $Z_3$ symmetry 
under which all the superfields have the same charge. This symmetry forbids 
the appearance of the usual bilinear $\mu$-term in the superpotential. The
$\mu$-term is generated spontaneously through the vacuum expectation value of
the singlet scalar $ \hat S$.    
In a similar way soft supersymmetry breaking potential can be written as 
\begin{eqnarray}
V_{\rm soft}= V_{\rm soft}^{\rm NMSSM}+V_{\it Singlet}\ , 
\label{softpot}
\end{eqnarray}
where $V_{\rm soft}^{\rm NMSSM}$ includes the MSSM soft supersymmetry breaking 
terms along with a few additional terms as shown below: 
\begin{eqnarray}
&&V_{\rm soft}^{\rm NMSSM}=V_{\rm soft}^{\rm MSSM}+
m_S^2 |S|^2 + \nonumber\\
&& \mbox{}\left(
A^H\lambda_H H_dH_u S 
%\right)\nonumber\\
+ A m \frac{\lambda_s}{3!} S^3 + \rm H.c
\right).
\end{eqnarray}
The term $V_{\it Singlet}$ is composed of the soft masses and the trilinear interactions 
corresponding to the fields $\tilde N_i$: 
\begin{eqnarray}
&&V_{\it Singlet}=
m^2_{{\tilde N}{\tilde N}^*} |\tilde N_i|^2 + \nonumber\\
&& \mbox{}\left(
A^\nu f^\nu_{ij}\tilde L_iH_u \tilde N_j
%\right)\nonumber\\
+ A m\frac{\lambda_{Ni}}{2} S \tilde N_i^2 + \rm H.c
\right).
\end{eqnarray}
We have taken a common trilinear coupling $A$ for the singlet fields $N_i$ and $S$
and $m$ is a mass scale. In a supergravity motivated scenario, it is a common practice
to choose $m = m_S = m_{{\tilde N}{\tilde N}^*}$ and also a universal trilinear parameter for the 
fields $S$, $\tilde N_i$. Since these fields are gauge singlet, we assume such
universality to hold also at the electroweak scale. Similarly, the mass
parameters $m_S$ and $m_{{\tilde N}{\tilde N}^*}$ are very much insensitive to 
Renormalization Group Equation (RGE) running and their values at the weak scale can 
be taken to be the same as the values at the high scale. In addition, we have chosen all 
the parameters $f^d_i$, $f^e_i$, ${\lambda_N}_i$, $\lambda_H$, $\lambda_s$,$f^u_{ij}$ and 
$f^\nu_{ij}$ to be real.

The scalar potential of this model can be written as 
\begin{eqnarray}
V = V_F + V_D + V_{\rm soft} \ , 
\label{Scalerpot}
\end{eqnarray}
where the neutral part of $V_F$ and $V_D$ can be written as 
\begin{widetext}
\begin{eqnarray}
V^{\it neutral}_F &=&
\sum_i\left|f^\nu_{ij}H^0_u \tilde N_j\right|^2 
+\left|\lambda_HH^0_uS\right|^2
+\left|f^\nu_{ij}\tilde \nu_i \tilde N_j+\lambda_HH^0_dS\right|^2 \nonumber\\
&&
+\sum_j\left|f^\nu_{ij}(\tilde \nu_iH^0_u)+{\lambda_N}_j \tilde N_jS\right|^2
+\left|\lambda_H(H^0_dH^0_u)+
%\sum_i
\frac{{\lambda_N}_i}{2}\tilde N_i^2
+\frac{\lambda_s}{2}S^2\right|^2 ,\label{vf} \\ 
\nonumber \\
V_D^{\it neutral} &=&
\frac{g_1^2+g_2^2}{8}\left(|H_u^0|^2-|H_d^0|^2
-\sum_i|\tilde{\nu}_i|^2\right)^2\ \label{vd}.
\end{eqnarray}
\end{widetext}
In the above, the repeated indices always mean to sum over the generations.
However, the summation sign is used in special cases if required.  
The VEVs are determined by the minimization of the potential 
(\emvide Eq.(\ref{Scalerpot}), (\ref{vf}) and (\ref{vd})). Here we explore the 
possibility when only scalar component of the gauge singlet superfield $\hat S$ 
acquires a VEV along with the doublet Higgs fields. The right-chiral sneutrino $\tilde N$ can 
only have a vanishing VEV and thus R-parity is unbroken. On the other hand, when
the right-chiral sneutrino ${\tilde N}$ acquires a VEV then R-parity is spontaneously broken 
and an effective bilinear R-parity violating term of the form $\epsilon_i L_iH_u$ is generated,
where $\epsilon_i \equiv f^\nu \langle \tilde N_i \rangle$. However, the case of spontaneous 
R-parity violation will be studied in a separate work\cite{debottam-future}.
Note that, a global continuous symmetry such as lepton number cannot be assigned to the 
superpotential involving the singlets $\hat S$ and $\hat N_i$. Thus this model is 
completely free from the unwanted Nambu-Goldstone boson even if the singlet 
scalar $S$ and/or {\bf $\tilde N_i$} acquire VEV. For more details the reader is referred 
to Ref.\cite{tev-seesaw,Kitano-2001}. 

Minimization of the scalar potential (\emvide Eq.(\ref{Scalerpot})) 
leads to the following conditions$\colon$ 
%%%%%%%%%%%%%%%%%%%%

\begin{widetext}
\begin{eqnarray}
{\partial V\over \partial v_d} &=&
2 v_d (m_{H_d}^2
        + \lambda_H^2 (v_u^2 + v_s^2)
        + {g_1^2 + g_2^2 \over 4}(v_d^2-v_u^2+\sum_i v_{\tilde \nu_i}^2) \\
&+& \tan\beta ( {1\over 2}\lambda_H\lambda_s v_s^2
+ {1\over 2}\lambda_H \lambda_{N_i} v_{\tilde N_i}^2 
+ A_H \lambda_H v_s ) )+ 2 \lambda_H f_\nu^{ij}v_sv_{\tilde \nu_i}
v_{\tilde N_j}
\nonumber ,\\
\nonumber
{\partial V\over \partial v_u} &=&
2 v_u (m_{H_u}^2
        + \lambda_H^2 (v_d^2 + v_s^2)
        - {g_1^2 + g_2^2 \over 4}(v_d^2-v_u^2+\sum_i v_{\tilde \nu_i}^2) \\
\nonumber
&+& f_\nu^{ij}f_\nu^{ik} v_{\tilde N_j} v_{\tilde N_k}
+  f_\nu^{ji}f_\nu^{ki} v_{\tilde \nu_j} v_{\tilde \nu_k} 
+ \cot\beta ( {1\over 2}\lambda_H\lambda_s v_s^2 
+ {1\over 2}\lambda_H \lambda_{N_i} v_{\tilde N_i}^2
+ A_H \lambda_H v_s ) ) \\
&+& 2 A_\nu f_\nu^{ij} v_{\tilde \nu_i} v_{\tilde N_j} 
+ 2 f_\nu^{ji}\lambda_{N_i}v_sv_{\tilde \nu_j}v_{\tilde N_i}\nonumber ,\\
\nonumber
{\partial V\over \partial v_s} &=&
2 v_s (m_S^2
            + \lambda_H^2(v_d^2+v_u^2)
            +\lambda_s \lambda_H v_d v_u 
            + \lambda_{N_i}^2 v_{\tilde N_i}^2+{1\over 2}A m\lambda_s v_s
%\\ \nonumber
%            &+& {1\over 2}A_S v_s 
            + {1\over 2}\lambda_s^2 v_s^2
+ {1\over 2}\lambda_s \lambda_{N_i}v_{\tilde N_i}^2)\nonumber\\ \nonumber
%&-&2\xi^3
&+& 2 A_H\lambda_H v_d v_u 
+ A m \lambda_{N_i} v_{\tilde N_i}^2
+ 2 f_\nu^{ij} v_{\tilde \nu_i}v_{\tilde N_j} (\lambda_H v_d 
+\lambda_{N_j} v_u )
\nonumber ,\\
\nonumber
{\partial V \over \partial v_{\tilde \nu_i}} &=&
2 v_{\tilde \nu_i} ({\tilde m}^2_i
+ {g_1^2 + g_2^2 \over 4}(v_d^2-v_u^2+\sum_j v_{\tilde \nu_j}^2) ) 
+ 2 A_\nu f_\nu^{ij} v_u v_{\tilde N_j} \nonumber
\\ \nonumber
&+& 2 \lambda_H f_\nu^{ij} v_d v_s v_{\tilde N_j}
+ 2  f_\nu^{ik}f_\nu^{jk} v_u^2 v_{\tilde \nu_j}
+ 2 f_\nu^{ij} \lambda_{N_j} v_u v_s v_{\tilde N_j}
+ 2f_\nu^{ij}f_\nu^{kl}v_{\tilde N_j}v_{\tilde \nu_k}v_{\tilde N_l}
\nonumber ,\\
\nonumber
{\partial V \over \partial v_{\tilde N_i}} &=&
2 v_{\tilde N_i} (m^2_{{\tilde N}{\tilde N}^*}
+ A m \lambda_{N_i} v_s
+ \lambda_{N_i} \lambda_H v_d v_u
+\frac{1}{2}\lambda_{N_i}\lambda_s v_s^2 + \lambda_{N_i}^2 v_s^2
+\frac{1}{2} \lambda_{N_i} \lambda_{N_j}v_{\tilde N_j}^2 )\\ \nonumber
&+& 2 f_\nu^{ji}f_\nu^{jk} v_u^2 v_{\tilde N_k}
+ 2 A_\nu f_\nu^{ji} v_u v_{\tilde \nu_j}
+ 2 \lambda_H f_\nu^{ji} v_d v_s v_{\tilde \nu_j}
+ 2 \lambda_{N_i} f_\nu^{ji} v_u v_s v_{\tilde \nu_j}
+ 2 f_\nu^{ji}f_\nu^{kl} v_{\tilde \nu_j} v_{\tilde N_l} v_{\tilde \nu_k}.
%\nonumber ,\\
\label{minimization}
\end{eqnarray}
\end{widetext}
Here $g_1$ and $g_2$ are the U(1) and SU(2) gauge couplings,
respectively, and $\tan\beta =v_u/v_d$. ${\tilde m}_i$ is the soft SUSY
breaking mass parameter of the left chiral sneutrinos. 
We have assumed that the neutral scalar fields can develop, in general, 
the following vacuum expectation values 
\begin{eqnarray}
v_d &=& \langle H^0_d \rangle;\hskip 0.2 cm   v_u = \langle H^0_u \rangle; 
\hskip 0.2 cm  v_s = \langle S \rangle; \hskip 0.2 cm \nonumber \\
v_{\tilde \nu_i}&=&\langle \tilde \nu_i \rangle; \hskip 0.2 cm
v_{\tilde N_i}= \langle \tilde N_i \rangle.
\end{eqnarray}
\nonumber
As has already been mentioned, in the present context we will consider the solutions $v_{\tilde N_i}
= v_{\tilde \nu_i} = 0$ and $v_s \neq 0$ to analyze the neutrino spectra. In our subsequent 
discussion, we will also ignore the terms in the minimization equations which are bilinear in the 
neutrino Yukawa couplings. Note that in order to generate very small masses for the active 
neutrinos ($\lsim$0.1 eV) using this TeV scale seesaw mechanism, the neutrino Yukawa couplings 
($f^\nu$) should be below ${\cal O}(10^{-6})$, which is around the magnitude of the electron Yukawa 
coupling.   

The VEV $v_s$ comes out as the solution of the following 
cubic equation (neglecting the Yukawa term), 
\begin{eqnarray}
%\nonumber
&& \hskip -0.6 cm\lambda_s^2 v_s^3 + Am\lambda_s v_s^2
+2v_s(m_s^2+\lambda_H^2v_u^2 + 
\lambda_H^2v_d^2+\lambda_H\lambda_sv_dv_u \nonumber\\ 
&+& \hskip -0.15 cm 
\lambda^2_{N_i}v_{\tilde N_i}^2+\lambda_s\lambda_{N_i}v_{\tilde N_i}^2)
%+\frac{1}{2}\lambda_n\lambda_sv_{\tilde \nu^c_3}v_s)
+2A_H\lambda_Hv_dv_u + Am\lambda_{N_i}v_{\tilde N_i}^2=0.\nonumber\\
\label{vs_soln}
\end{eqnarray}
%\end{widetext}
The solutions of the foregoing equation involve soft parameters $Am$, 
$A_H$ and $m_s^2$. In fact these parameters cannot be much away from TeV 
values to have $v_s$ $\sim$ TeV. In particular, the soft parameter 
$A_H$ and $Am$ are crucial to produce non zero VEV for the field $S$. Any consistent 
solution that yields $v_s\ne0$ but $v_{\tilde N_i}=0$ requires $|A|\ge3$ and 
also $\lambda_H \le 1$, $m \ge 100$ GeV, $m_S \ge 100$ GeV\cite{Kitano-2001}.
Similarly we also choose the couplings $\lambda_s$,
$\lambda_{N_i}$ in such a manner so that the condition for global minima is
always satisfied.

\section{\small {Neutrino masses and mixing: R-parity conserving NMSSM}}
Let us now discuss in detail the generation of neutrino masses and mixing in this model. 
Note that this model is different from the models where MSSM is extended with three 
right-handed singlet neutrino superfields. This is because in those models the right handed 
neutrino mass scale is not tied up with the electroweak symmetry breaking scale and is 
assumed to be very high ($\sim 10^{15}$ GeV or so).  

\subsection{Seesaw masses}

At the tree level, the $(3 \times 3)$ light neutrino mass matrix, 
that  arises via the seesaw mechanism has a very well-known structure 
given by 
\begin{equation}
m_{\nu}^{tree} = -m_DM^{-1}_Rm^T_D,
\label{seesaw-formula}
\end{equation}
%%%%%%%%%%%%%%%%%%%%%%%%%%%%%%%%%%%%%%%%%%%%%%%%%%%%%%%%%%%%%%
where $m_D$ represents the lepton number conserving $(3 \times 3)$ `Dirac' 
mass matrix and $M_R$ represents the lepton number violating 
$(3 \times 3)$ `Majorana' mass matrix. Note that, after the EWSB, when the
scalar component of ${\hat S}$ gets a VEV, in the effective Lagrangian
we can assign a lepton number -1 for the fields $N_i^c$ and ${\tilde N}_i$ (contained
in the superfield ${\hat N}_i$). The relevant part of the effective Lagrangian which 
encompasses both neutrino and sneutrino fields is given by
\begin{eqnarray}
 -{\cal L}_{eff} &=& \frac{1}{2}(\lambda_{Ni} v_s) {N^c_i}{N^c_i} + 
f^\nu_{ij}{\nu_i}{v_u}{N^c_j} +\rm H.c. \nonumber \\
&+& m^2_{\tilde\nu_i} {\tilde \nu_i} {\tilde \nu^{\star}_i} +
(m_{\tilde N_i {\tilde N_i}^*}^2+ \lambda_{N_i}^2 v_s^2){\tilde N_i} 
{\tilde N^{\star}_i} \nonumber \\
 &+& (B^\nu_{ij} {\tilde{\nu}_i}{\tilde N_j}+
{B^{{\prime}\nu}_{ij}} {\tilde{\nu}_i}{\tilde N^{\star}_j}+
B_{Ri} {\tilde N_i}{\tilde N_i}
 + {\rm H.c.}), \nonumber \\
\label{eq-eff-lagrangian}
\end{eqnarray}
where the coefficients have the following meaning
\begin{eqnarray}
m^2_{\tilde\nu_i} &=& {\tilde m}^2_i+ \frac{1}{2}m_Z^2 cos2\beta, 
\nonumber \\
B^\nu_{ij} &=& A^\nu f^\nu_{ij}{v_u}+\lambda_Hf^\nu_{ij}v_dv_s, \nonumber \\
B^{{\prime}\nu}_{ij}&=&f^\nu_{ij} \lambda_{Nj}{v_u}{v_s},\nonumber \\
B_{Ri}&=&\frac{1}{2}({\lambda_H\lambda_{N_i}v_dv_u}
+\frac{\lambda_s\lambda_{N_i}v_s^2}{2}
+ Am \lambda_{Ni}{v_s}).  \nonumber \\
\label{loopcoeff1}
\end{eqnarray}
It is easy to see from Eq.(\ref{eq-eff-lagrangian}) that 
${m_D}_{ij}\equiv f^\nu_{ij}v_u $ and $M_R{_i}=\lambda_{Ni} v_s$, 
which in turn provide neutrino masses at the tree level through
Eq.(\ref{seesaw-formula}). Note that, in Eq.(\ref{loopcoeff1}) we have
neglected a term $\sim m^2_D$ in the expression for $m^2_{\tilde\nu_i}$
since it is much smaller compared to the other terms.

\begin{figure}
\begin{center}
\includegraphics[height=1.24in,width=3in]{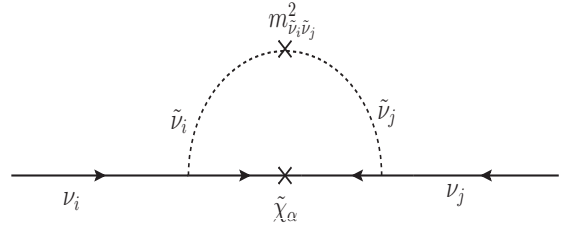}
  \caption{One-loop contribution to $m_{\nu}$ when $v_{\tilde N_i}= v_{\tilde \nu_i}=0$. 
Here $m^2_{\tilde \nu_i \tilde \nu_j}$ represents the sneutrino ``Majorana" mass term 
which generates the neutrino mass involving the sneutrino-neutralino loop.}
  \label{looprpc}
  \end{center}
\end{figure}

The tree level neutrino masses may receive dominant radiative 
corrections at the one-loop level. It has been shown in models of
MSSM with right-handed neutrino superfields, that the loop 
contributions can be as large as the tree level value, though 
the result depends on the soft SUSY breaking parameters 
\cite{grossman-haber,davidson-king}. In $R$-parity conserving 
scenarios the leading contribution to neutrino masses at the 
one-loop level arise from $\Delta L=2$ terms in the sneutrino
sector.  These bilinear interaction terms involving the heavy
right-handed sneutrinos fields ${\tilde N}_i$ are 
${B^{{\prime}\nu}_{ij}} {\tilde{\nu}_i}{\tilde N^{\star}_j}$,
and $B_{Ri} {\tilde N_i}{\tilde N_i}$ as can be seen from 
Eq.(\ref{eq-eff-lagrangian}). In association with the 
$\Delta L=0$ term i.e.,  $B^\nu_{ij} {\tilde{\nu}_i}{\tilde N_j}$ 
these $\Delta L=2$ terms generate lepton number violating 
``Majorana" like mass terms ($m^2_{{\tilde \nu} {\tilde \nu}}
{\tilde \nu} {\tilde \nu} + h.c.$) for the left-handed sneutrinos. 
In fact, this can be seen as a scalar seesaw analogue of the usual 
fermionic seesaw mechanism to generate small masses for the light 
active neutrinos\cite{davidson-king}. This effective Majorana sneutrino 
mass term in turn induces one-loop radiative corrections to neutrino Majorana 
masses via the self-energy diagram as shown in Fig.\ref{looprpc}. 
However, rather than computing the one-loop contribution to neutrino masses 
using the above method, we would choose a different but more general
procedure as explained below.
\begin{figure}
\begin{center}
\includegraphics[height=1.24in,width=3in]{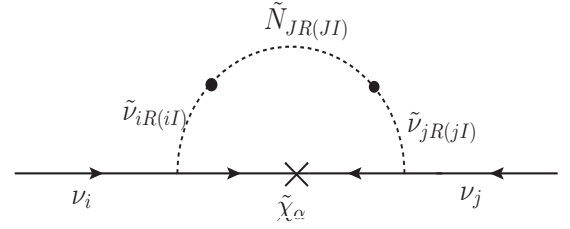}
  \caption{The same one-loop contribution to $m_{\nu}$ as in Fig.\ref{looprpc} but 
represented in a different way. Here ${\tilde N}_{JR}$ and ${\tilde N}_{JI}$ are 
right handed sneutrino mass eigenstates which couple to ${\tilde \nu}_{i,j}$ to 
produce the one-loop effective neutrino mass term.}
\label{looprpc1}
  \end{center}
\end{figure}

We begin by decomposing the sneutrino fields in terms of real and imaginary 
components. Thus one has 
\begin{eqnarray}
\tilde \nu _{i} &=& {\tilde \nu _{iR} +i
\tilde \nu _{iI} \over \sqrt 2 },\hskip 0.3 cm \tilde N_{i} = 
{\tilde N
_{iR} +i\tilde N_{iI} \over \sqrt 2 },\label{decompose}
\end{eqnarray}
where the components 
$\tilde \nu _{iR}, \tilde N_{iR}$ are the $CP$-even and 
$\tilde \nu _{iI}, \tilde N_{iI}$ are the $CP$-odd scalar fields. 
The mass terms of these scalars may be evaluated using the definition
%\begin{widetext}
\begin{eqnarray} 
M^{2}_{R,ij} & = & {\dh^2 V \over \dh
\Phi_{i R} \dh \Phi_{j R} } ,\ M^{2}_{p,ij} = {\dh^2 V \over \dh
\Phi_{iI} \dh \Phi_{j I} },  
\end{eqnarray}
where $\Phi$ represents a generic scalar field.
Accordingly one obtains the following diagonal mass terms 
(assuming the right-chiral sneutrino states to be flavor diagonal) 
for the  $CP$-even and $CP$-odd right-chiral sneutrinos: 
%\begin{widetext}
\begin{eqnarray} 
M^2 _{R, \tilde N_i \tilde N_i}&=& m^2_{\tilde N_i \tilde N_i^\star} + \lambda^2_{Ni} v_s^2
\nonumber \\
&+& \hskip -0.25 cm 
(\lambda_H \lambda_{Ni} v_d v_u + 
\frac{1}{2} \lambda_{Ni} \lambda_s v_s^2 +A m \lambda_{Ni} v_s) 
\nonumber \\
M^2 _{P, \tilde N_i \tilde N_i}&=&  m^2_{\tilde N_i \tilde N_i^\star} 
+ \lambda_{Ni}^2 v_s^2  
\nonumber \\ &-& \hskip -0.25 cm 
(\lambda_H\lambda_{Ni}v_dv_u+
\frac{1}{2}\lambda_{Ni}\lambda_sv_s^2 +Am\lambda_{Ni}v_s). \nonumber \\ 
\label{mrmp}
\end{eqnarray}
Similarly, the interactions between $\tilde N_{iR,(iI)}$ and 
$\tilde \nu_{iR,(iI)}$ read as   
\begin{eqnarray} 
C _{R,\tilde \nu _{i} \tilde N _{j} }&=&  f^\nu_{ij}\lambda_H v_d v_s 
+f^\nu_{ij}\lambda_{Nj} v_u v_s + A^\nu f^\nu_{ij} v_u,\nonumber
\\ 
C_{P,\tilde \nu _{i} \tilde N _{j} }&=&  -f^\nu_{ij}\lambda_H v_d v_s 
+f^\nu_{ij}\lambda_{Nj} v_u v_s - A^\nu f^\nu_{ij} v_u. \nonumber \\
%\nonumber \\ 
\label{mixing_sneu}
\end{eqnarray}
%\end{widetext}
The diagonal left-chiral sneutrino mass terms are shown in 
Eq.(\ref{loopcoeff1}). As we can see, the off-diagonal terms involving the
left-chiral and right-chiral sneutrinos are much smaller compared to the 
diagonal terms since they are proportional to the small neutrino Yukawa couplings
($f^\nu \sim 10^{-6}$). Hence, we can compute the one-loop correction to the neutrino 
mass due to the small mixing of the right-chiral sneutrinos with the left-chiral 
sneutrinos. This is shown in Fig.\ref{looprpc1}.
Note that the right-chiral sneutrino mass matrix contains bilinear 
terms like $\lambda_H\lambda_{N_i}v_dv_u$,
$\frac{\lambda_s\lambda_{N_i}v_s^2} {2}$ which are
originated from the F-term contribution in the scalar potential. These
are the new contributions to the right sneutrino masses in the present model 
and thus they are absent in seesaw models of MSSM with only right handed neutrino superfields. 
These terms will have important roles to play while calculating the one-loop correction
to the neutrino mass matrix, even when the relevant soft breaking trilinear parameters are smaller. 
The loop contribution can be written as,
\begin{widetext}
\begin{eqnarray} 
(m_\nu )_{ij} ^{loop} & = &
{g_2^2\over 4 } \sum _\alpha m_{\wtchi _\alpha } (N_{\alpha5} - \tan 
\theta_w N_{\alpha4} ) ^2
[ \sum _{J=1,2,3}C_{R_{iJ}}C_{R_{jJ}} I_{4} ( m _{\tilde \nu _{iR}}, 
m _{\tilde
\nu _{jR}} m_{\wtchi _\alpha }, M_{\tilde N_{JR}} ) \nonumber \\ &-&  
\sum _{J=1,2,3} C_{P_{iJ}}C_{P_{jJ}}
 I_{4} ( m _{\tilde \nu _{iI}}, m _{\tilde \nu _{jI}}, m_{\wtchi _\alpha
}, M _{\tilde N_{JI}} ) ],
\label{looprpcbb}
\end{eqnarray}
%\end{widetext}
where the integral $I_4$ is given by,
%\begin{widetext}
\begin{eqnarray}
I_{4} ( m_{\tilde \nu _{iR}}, m
_{\tilde \nu _{jR}}, m_{\wtchi_\alpha }, m_{X_J} ) & = & \int {d^4 q\over i 
(2\pi )^4 } {1\over (q^2
- m^2_{\tilde \nu _{iR} } )(q^2 - m^2_{\tilde \nu _{jR} } ) (q^2 -
m_{\wtchi_\alpha } ^2) (q ^2 - m^2 _{X_J})}.
\end{eqnarray}
Here $X_J$ denotes right chiral sneutrino states $\tilde N_{JR}$ or 
$\tilde N_{JI}$. One can always evaluate $I_4$ with the following analytical 
expressions,
\begin{eqnarray} 
I_4 (m_1, m_2,m_3,m_4) & = & {1
\over m_3 ^2 - m_4 ^2} [ I_3 (m_1, m_2,m_4)- I_3 (m_1, m_2,m_3) ] ,\cr
I_3 (m_1, m_2,m_3) & = & {1 \over m_2^2- m_3^2} [I_2 (m_1, m_2) - I_2
(m_1, m_3)] , \cr I_2 (m_1, m_2) & = & {1\over (4\pi )^2} { m_2 ^2
\over m_1 ^2 - m_2^2 } \log {m_1^2 \over m_2^2}. 
\end{eqnarray}
Here, $m_{\wtchi _\alpha }$ represents the eigenvalues of the 
NMSSM neutralino mass matrix. 
In the weak interaction basis $\left(\wt S,{\wt H_d^0},{\wt H_u^0},\widetilde{B}^0,
\widetilde{W}^0\right)$, the mass matrix can be written as 
\begin{eqnarray}
{\cal M}&=&\left(
\begin{array}{ccccc}
\lambda_Sv_s & \lambda_Hv_u & \lambda_Hv_d & 0 & 0\\
\lambda_H v_u & 0 &\lambda_Hv_s & -g_1v_d/\sqrt{2} & 
g_2v_d/\sqrt{2}\\
\lambda_H v_d & \lambda_Hv_s & 0 & g_1v_u/\sqrt{2} & 
-g_2v_u/\sqrt{2}\\
0 &  -g_1v_d/\sqrt{2} & g_1v_u/\sqrt{2} & M_1 & 0\\
0 &  g_2v_d/\sqrt{2} & -g_2v_u/\sqrt{2} & 0 & M_2
\end{array}\right) \ .
\label{nmssm-matrix}
\end{eqnarray}
\end{widetext}
The mixing matrix elements $N_{\alpha5}$ and $N_{\alpha4}$ are the wino and bino
component of the neutralino ${\tilde \chi}_\alpha$. The expression 
(\emvide Eq.(\ref{looprpcbb})) is the most general to compute 
the one-loop diagram (\emvide Fig.\ref{looprpc1}). Nevertheless, we would consider 
a simplified scenario for illustration. In particular, we assume (i) identical
values of $\lambda_{Ni}$ ( $\lambda_{Ni} \equiv \lambda_N$) for all 
three generations and (ii) soft-masses of the sneutrinos (both $\tilde \nu_i$ and 
$\tilde N_i$) are flavor blind. This results into identical mass values for all
three $CP$-even right chiral sneutrinos ($M_{\tilde N_{RJ}}\equiv M_{\tilde N_{R}}$) 
and also for the three $CP$-odd states ($M_{\tilde N_{IJ}} \equiv M_{\tilde N_{I}}$).
With these assumptions, it is possible to factor out the 
flavor structure from Eq.(\ref{looprpcbb}) and denote the remaining as the loop factor (LF) 
which is merely a constant.
Then the loop contribution can be cast into a convenient form given by
\beq
(m_\nu^{\rm loop})_{ij}=(LF)\sum_{k=1}^3{{f^\nu_{iJ}}{f^\nu_{jJ}}},
\label{mnuloop_simplified}
\eeq
where
\begin{eqnarray} 
LF &=&
{g_2^2\over 4 } \sum _\alpha m_{\wtchi _\alpha } (N_{\alpha5} - \tan 
\theta_w N_{\alpha4} ) ^2 \nonumber \\ && \hskip -0.9 cm
( I_{4} ( m _{\tilde \nu }, 
m _{\tilde\nu }, m_{\wtchi _\alpha }, M _{\tilde N_{R}}) C_R^2  
-  I_{4} ( m _{\tilde \nu}, m _{\tilde \nu}, m_{\wtchi _\alpha
}, M_{\tilde N_{I}} ) C_P^2 ).\nonumber \label{lfactor}
\\
\end{eqnarray}
Here $C_R$ and $C_P$ represent the coefficients of $f^\nu_{ij}$ in Eq. 
(\ref{mixing_sneu}) and given as 
\begin{eqnarray} 
C _{R}&=& \lambda_H v_d v_s
+\lambda_{N} v_u v_s + A^\nu v_u,
 \\ 
C_{P}&=&  -\lambda_H v_d v_s 
+\lambda_{N} v_u v_s - A^\nu v_u.
%\nonumber \\ 
\end{eqnarray}
Let us note that the coefficient $B_{Ri}$ can be written as 
\begin{eqnarray}
B_{Ri} = B_N M_R
\end{eqnarray}
where 
\begin{eqnarray}
\label{exprn-bn-mn}
B_N &=& \frac{1}{2}(\lambda_H v_d v_u/v_s + \frac{\lambda_s v_s}{2}
+ Am), \nonumber \\
M_R &=& \lambda_N v_s.
\end{eqnarray}

Consequently the one-loop contribution can be cast into the well known
form\cite{davidson-king,grossman-haber}
\beqn 
\label{loophabermass1}
{m_\nu^{(loop)}}_{ij} &=& -{g_2^2{\Delta m_{{\tilde \nu}}}_{ij} \over 32 \pi^2 \cos^2 \theta_W}
\sum_\alpha \,f(y_\alpha) |N_{\alpha k}|^2\,,\\ \nonumber
f(y_\alpha) &=& {\sqrt{y_\alpha}\left[y_\alpha-1-\ln(y_\alpha)\right]\over 
(1-y_\alpha)^2},
\eeqn
where $y_\alpha\equiv{m_{\tilde \nu}^2/m_{\tilde\chi^0_\alpha}^2}$ and
$N_{\alpha k}\equiv N_{\alpha 5}\cos\theta_W-N_{\alpha 4}\sin\theta_W$ 
is the neutralino mixing matrix element and to order in $1/M^3_R$ the left 
sneutrino mass difference relative to the light neutrino mass is given by
\beqn 
{{\Delta m_{\tilde \nu}}_{ij}  \over  {m_\nu}_{ij}} &\simeq&
{{2 (A_\nu+\mu \cot\beta-B_N - \frac{B_N (A_\nu+\mu \cot\beta)^2}{M^2_R})} \over m_{\tilde \nu}}.\nonumber\\
\label{sneutrino-mass-splitting}
\eeqn
Here we have used the relation $\Delta m^2_{\tilde \nu} = 2m_{\tilde \nu} 
\Delta m_{\tilde \nu}$ and $m_{\tilde \nu}$ is an average left-sneutrino mass.
In the present case all left handed sneutrino soft masses are assumed to be 
identical. The sneutrino Majorana mass $m^2_{{\tilde \nu}{\tilde \nu}}$ shown in Fig.\ref{looprpc}
is related to $\Delta m^2_{\tilde \nu}$ as $m^2_{{\tilde \nu}{\tilde \nu}} = 
\frac{1}{4}\Delta m^2_{\tilde \nu}$\cite{davidson-king}. The quantity $\mu$ is defined as
$\mu = \lambda_H v_s$.

In order to reproduce the result in Eq.(\ref{loophabermass1}), we assumed that 
$B_N, m_{\tilde N \tilde N*} < M_R$ and $A_\nu > B_N$. Now, in addition if we assume
$M_R > A_\nu$, the last term becomes negligible compared to the other terms
in the expression Eq.(\ref{sneutrino-mass-splitting}) and this keeps only
the terms to leading order in $1/M_R$. However, this is not 
always true as all soft SUSY breaking mass parameters as well as the right handed neutrino
masses may have similar magnitudes as in the present scenario. Hence, rather than 
using Eq.(\ref{loophabermass1}), we evaluate the neutrino 
mass terms corrected up to one loop order, from
\beq
(m_\nu^{\rm total})_{ij}=(\frac{-v^2_u}{M_R}+LF)\sum_{k=1}^3{{f^\nu_{iJ}}{f^\nu_{jJ}}}.
\label{mnutotal_simplified}
\eeq
Clearly, the coefficient of the loop contribution shifts the tree level neutrino masses
by a constant amount. This coefficient involves the soft SUSY breaking parameters and
in this work we explore the effect of these parameters on the neutrino mass matrix.

This simple structure of the neutrino mass matrix 
(\emvide Eq.(\ref{mnutotal_simplified})) can indeed be very helpful to examine the 
neutrino mixing pattern. In particular, we are interested to explore the conditions
which could yield the mixing matrix into a tri-bimaximal structure. Thus we 
compare Eq.(\ref{mnutotal_simplified}) with Eq.(\ref{neutribi}), where the latter 
provides with the neutrino mass matrix consistent with the tri-bimaximal mixing pattern. 
Then, with a symmetric neutrino Yukawa matrix, neutrino masses can be evaluated using the 
following expressions:

%\begin{widetext}
\begin{eqnarray}
\frac{2}{3}m_1+\frac{1}{3}m_2 &=& C[(f^\nu_{11})^2
+(f^\nu_{12})^2+(f^\nu_{13})^2],
\nonumber	 \\
\frac{1}{6}(m_1+2m_2+3m_3) &=&C[(f^\nu_{12})^2+
(f^\nu_{22})^2+(f^\nu_{23})^2] ,
\nonumber	 \\
&=& C[(f^\nu_{13})^2+(f^\nu_{23})^2+(f^\nu_{33})^2],
\nonumber	 \\
\frac{1}{3}(-m_1+m_2) &=&C[f^\nu_{11}f^\nu_{12}+f^\nu_{12}
f^\nu_{22}+f^\nu_{13}f^\nu_{23}] ,
\nonumber	 \\
\frac{1}{3}(m_1-m_2) &=&C[f^\nu_{11}f^\nu_{13}+f^\nu_{12}
f^\nu_{23}+f^\nu_{13}f^\nu_{33}] ,
\nonumber	 \\
\frac{1}{6}(-m_1-2m_2+3m_3) &=&C[f^\nu_{12}f^\nu_{13}+f^\nu_{22}
f^\nu_{23}+f^\nu_{23}f^\nu_{33}]. \nonumber	 \\
\end{eqnarray}
%\end{widetext}
Here  the constant $C$ is defined as $(\frac{-v^2_u}{M_R}+LF)$. As a simple
choice we consider, $f^\nu_{22}=f^\nu_{33}$ and also 
$f^\nu_{12}=f^\nu_{13}=0$ to obtain the solutions.  This choice, 
coupled with the consistency condition $f^\nu_{11}=f^\nu_{22}-f^\nu_{23}$, 
leads to the following solutions of the neutrino spectra
\begin{eqnarray}
m_1&=& m_2=(\frac{-v^2_u}{M_R}+LF)(f^\nu_{11})^2, \nonumber \\  
m_3&=& (\frac{ -v^2_u}{M_R}+LF)(2 f^\nu_{22}- f^\nu_{11})^2.
\label{neumass}
\end{eqnarray}
It is obvious that the mass pattern as depicted above satisfies the desired 
tri-bimaximal structure of the neutrino mixing. The mass terms as expected,
contain tree level contributions which are always negative. On the other hand, 
the loop contribution can go both ways depending on the sign of the soft SUSY breaking 
parameters. For a large $B_{R}$, which primarily depends on $Am$, the 
radiative correction to the neutrino masses could be enhanced to supersede the tree level 
results\cite{grossman-haber}. 

Before presenting the numerical results a few comments regarding the lepton flavor
violating (LFV) processes are in order. Recall that we assume flavor diagonal mass terms for
the left and right chiral sneutrinos. The loop induced processes like $\mu \rightarrow e \gamma$,
$\tau \rightarrow e \gamma$ or $\tau \rightarrow \mu \gamma$ can get contributions primarily
via the couplings $B^\nu_{ij}$ or $B^{\prime\nu}_{ij}$ (see Eq.(\ref{eq-eff-lagrangian}) and 
(\ref{loopcoeff1})). Clearly, any such contribution at the leading order would involve a product of two 
small neutrino Yukawa couplings $f^\nu_{ij}$ and are expected to be very suppressed. Moreover, our assumption 
$f^\nu_{12}=f^\nu_{13}=0$ would lead to vanishing contributions for the processes $\mu \rightarrow e \gamma$ 
and $\tau \rightarrow e \gamma$ in this model. 
 
We now explore whether the obtained mass pattern could fit with
the different hierarchical structure that we know so far. In particular, 
we show our numerical results to identify the regions in the parameter
space consistent with the normal, inverted and quasi-degenerate neutrino mass 
pattern. In the numerical computation we choose different soft parameters 
and couplings in such a way, that the proper minima condition of the scalar potential is 
always satisfied\cite{Kitano-2001}.

%%%%%%%%%%%%%%%%%%%%%%%%%%%%%%%%%%%%%%%%%%%%%%%%%%%%%%%%%%%%% 
The choices of various parameters are listed below.
The value of $\tan\beta$ is taken to be equal to 10. In addition to that, other
parameter choices are \\ 
(I) Superpotential parameters:
$\lambda_h=-0.3$, \hskip 0.2 cm $\lambda_s=0.6$, 
\hskip 0.2 cm $\lambda_{N1}=\lambda_{N2}=\lambda_{N3}=\lambda_{N}=0.2$, \\ 
and \\
\noindent (II) Soft SUSY breaking parameters: 
$m_S$ = 100 GeV, $m_{{\tilde N}_i{\tilde N}_i^*}$ = 300 GeV,   
$m_{{\tilde \nu}_i}$ = 100 GeV, 
$A_H$ = 100 GeV, 
$A_\nu$ = 1000 GeV.\\

Apart from the above parameters which are fixed to the quoted values, 
we have also varied the parameter $Am$ in the calculation. This would 
cause changes in $v_s$ (\emvide Eq.(\ref{vs_soln})), 
which in turn produces variation in the neutrino
spectrum. We list the values of $Am$ and $v_s$ in table I.

\begin{table}[!ht]
\begin{center}\
\begin{tabular}{|c|c|c|c|c|}
\hline
$Am$ (GeV) &-600.0  &-800.0 &-1000.0&-1200.0\\
\hline
$v_s$ (GeV) &927.56  &1280.76 &1625.18&1965.67\\
\hline
\end{tabular}
\end{center}
   \caption{Different values of $v_s$ corresponding to the
different values of the coupling parameter $Am$.}
\label{tablevs}
\end{table}

\subsection{Different Neutrino Spectra:}
The two mass-squared differences shown in Eq.(\ref{eq:data}) 
indicate three possible neutrino mass hierarchies\cite{hierarchy}, namely
\begin{enumerate}
\item {Normal Hierarchy:}
this neutrino mass pattern can be established
if $m_1,m_2$ and $m_3$ are related with the observables 
$\sqrt{\Delta m^2_{21}}$ and $\sqrt{|\Delta m^2_{32}|}$ as
\begin{eqnarray}
m_1\approx m_2\sim \sqrt{\Delta m^2_{21}},\qquad
m_3\sim \sqrt{|\Delta m^2_{32}|}.
\label{nor-cond}
\end{eqnarray}
However, in principle $m_1$ can also be much smaller than $m_2$ or even be
zero. Since in this case $m_3$ is much greater than both $m_1$ and $m_2$, we
can approximately use the relation shown in Eq.(\ref{nor-cond}) for illustration. 
 
\item {Inverted Hierarchy:}
this hierarchical scenario can be achieved if one chooses
\begin{eqnarray}
m_1\approx m_2\sim \sqrt{|\Delta m^2_{32}|},\qquad
m_3\ll \sqrt{|\Delta m^2_{32}|}.
\label{inv-cond}
\end{eqnarray}
We assume the maximum possible value for $m_3$ to be $\sim 0.01$ eV while the 
minimum value could be vanishing. Obviously, the solar mass squared difference
$\Delta m^2_{21}$ will come from the small mass splitting between $m_2$ and $m_1$, where
$\Delta m^2_{21} \ll m_2, m_1$. Hence, for a simple minded analysis we can assume that 
$m_2 = m_1$.
\item {Degenerate Masses: }
finally this scenario is defined by 
\begin{eqnarray}
m_1\approx m_2\approx m_3\gg \sqrt{|\Delta m^2_{32}|},
\label{deg-cond}
\end{eqnarray}
Here we assume that the upper bound of the neutrino masses could be $0.33$ eV, which
comes from the cosmological observations. The lower bound 
is chosen to be $0.1$ eV.
\end{enumerate}

\begin{figure}
%\begin{center}
\subfigure[]{\label{fig:nor_hie-a}\includegraphics[height=3in,width=3in]{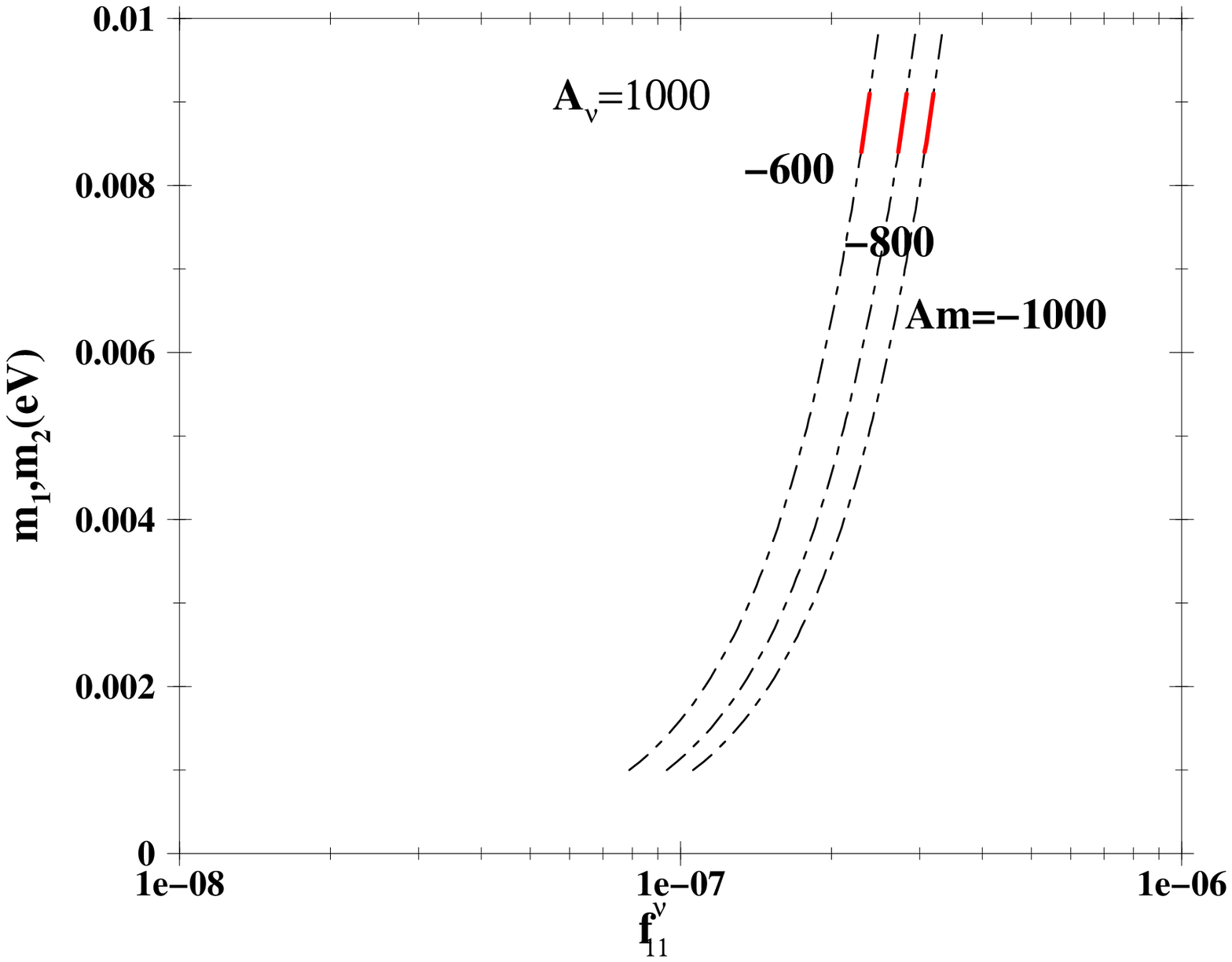}}
\subfigure[]{\label{fig:nor_hie-b}\includegraphics[height=3in,width=3in]{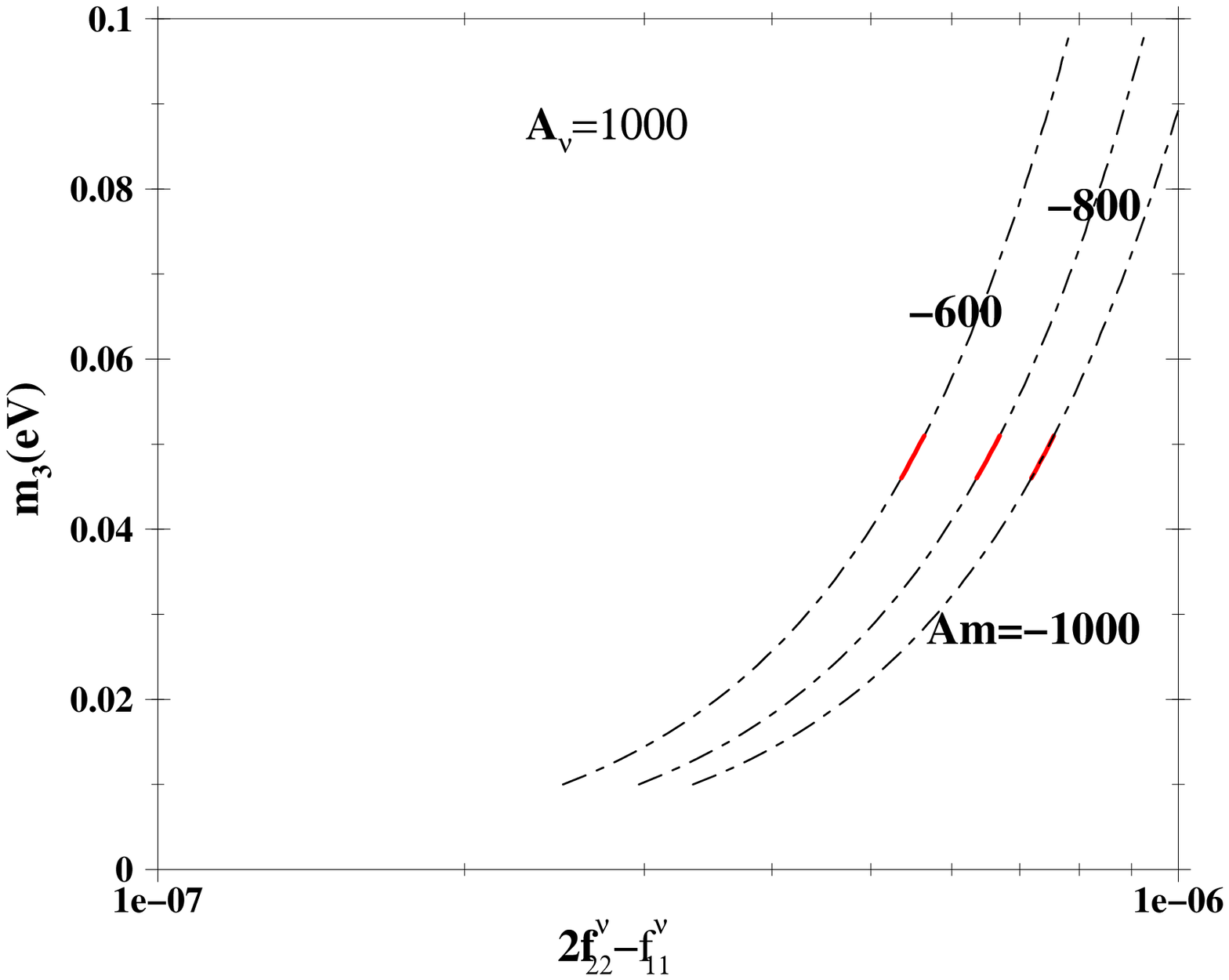}}
\caption{Normal hierarchy: variation of $m_{\nu}$ with the Yukawa parameters.
The red (solid) segments denote the range of the Yukawa parameters that satisfy
neutrino data. Each contour represents a separate set of $v_s$ and $Am$,
as given in Table.\ref{tablevs}. All mass parameters are in GeV.}
\label{fig:nor_hie}
\end{figure}

\begin{figure}
\subfigure[]{\label{fig:inv_hie-a}\includegraphics[height=3in,width=3in]{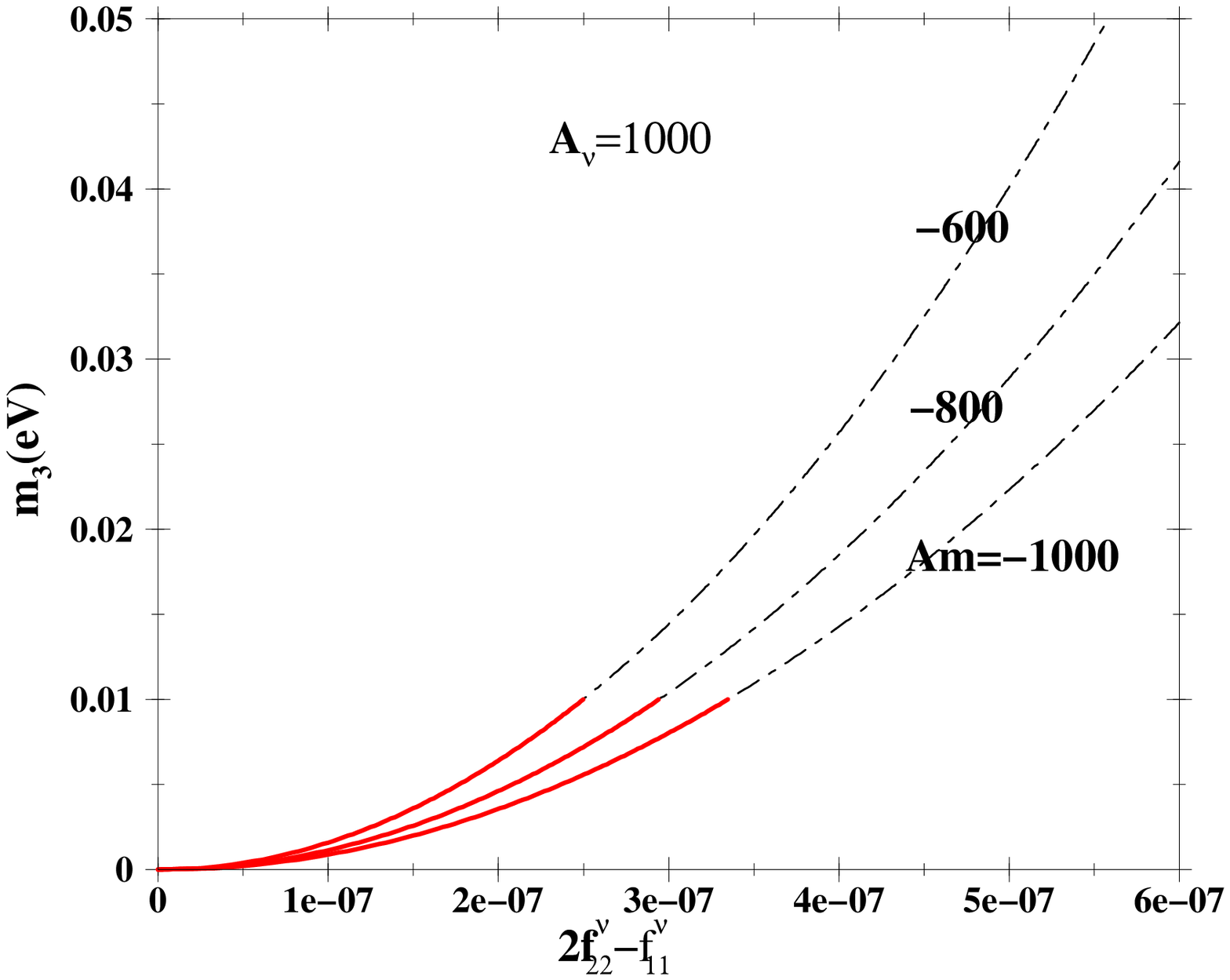}}
\subfigure[]{\label{fig:inv_hie-b}\includegraphics[height=3in,width=3in]{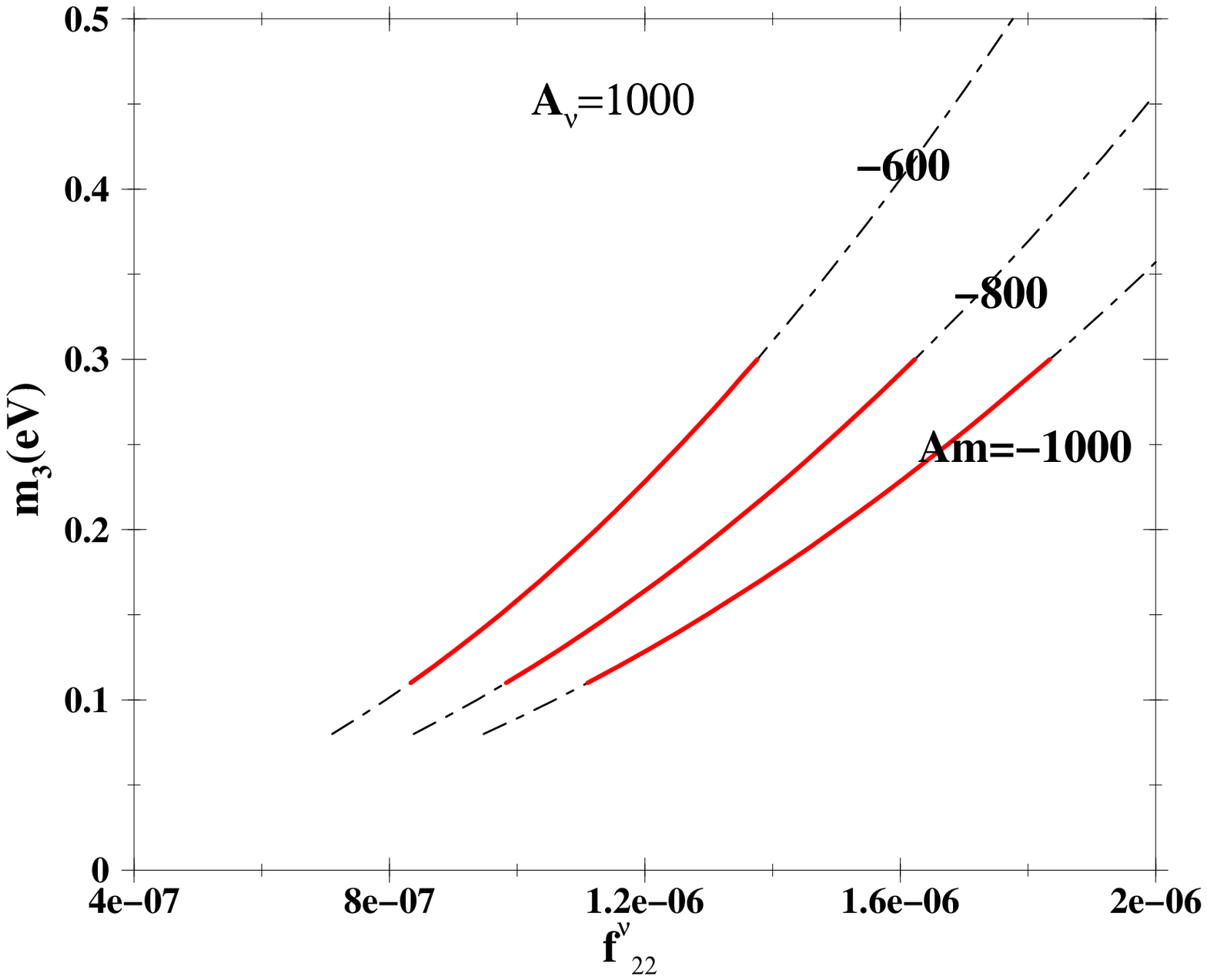}}
\caption{Variation of $m_3$ with the corresponding Yukawa parameter is 
shown for (a) inverted hierarchical mass pattern and also for (b) the degenerate spectrum. 
All mass parameters are in GeV.}
\label{fig:inv_hie}
\end{figure}

In Fig\ref{fig:nor_hie}, three neutrino mass eigenvalues $m_1,m_2,m_3$, 
consistent with the normal hierarchical pattern, are plotted as functions of neutrino 
Yukawa couplings. The difference in the contours manifests how the neutrino masses depend on 
the soft bilinear coupling parameter ($Am$). The variation occurs, as $v_s$ depends on ($Am$), 
thereby acquiring a different value at the global minima which has already been 
mentioned in Table~\ref{tablevs}. In particular $|v_s|$ always increases as 
we increase $|Am|$ parameter which in turn increases the right handed neutrino masses. 
This results into a smaller value for $m^{tree}_\nu$. On the other hand, loop correction does not
increase appreciably by this small variation of $Am$ if $A_\nu$ is around TeV scale as we will 
discuss later. We should note here that neutrino loop correction is always 
an order of 
magnitude smaller compared to the tree level value for the parameters we have chosen. Thus with 
increase in $Am$ parameter, one requires large values of Yukawa couplings to satisfy the neutrino 
data. The red zone in each contour (\emvide Fig\ref{fig:nor_hie}) represents the range of the Yukawa 
couplings that can satisfy the neutrino data.

In case of inverted hierarchy, we have shown the variation of $m_3$ 
with the respective Yukawa couplings in Fig.\ref{fig:inv_hie-a}.
The other mass parameters $m_1,m_2$ depend on the 
Yukawa coupling $f^\nu_{11}$, but that can be estimated 
from the Fig.\ref{fig:nor_hie-b} if in that plot we replace $m_3$ in the y-axis by 
$m_1/m_2$ and $2f^\nu_{22}-f^\nu_{11}$ in the x-axis by $f^\nu_{11}$ 
(\emvide Eq.(\ref{neumass})). In fact knowing the value of the Yukawa coupling 
$f^\nu_{11}$ would allow us to determine the coupling $f^\nu_{22}$.

The Fig.\ref{fig:inv_hie-b} depicts the variation of $m_3$  
with the Yukawa coupling $f^\nu_{22}$ for quasi-degenerate mass scenario. In this scenario, 
the neutrino spectrum is approximately degenerate i.e., $m_1$, $m_2$ and $m_3$ turn out to be 
almost identical if one chooses $f^\nu_{23}$ much smaller compared to the diagonal Yukawa coupling 
$f^\nu_{22}$, which essentially means that $f^\nu_{22} \approx f^\nu_{11}$. 

\begin{figure}
\includegraphics[height=3in,width=3in]{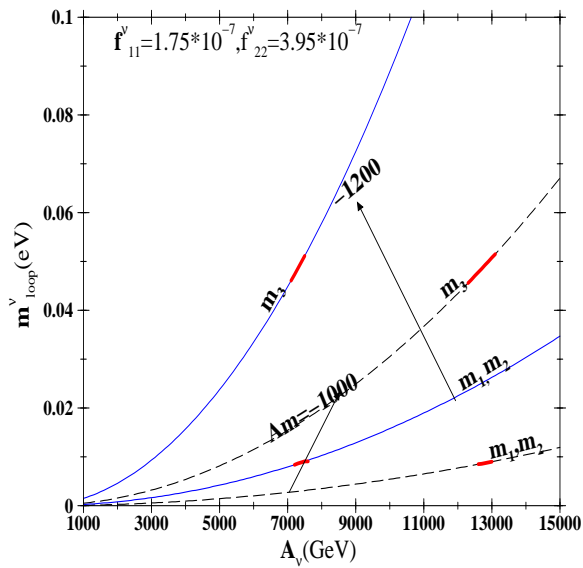}
\caption{Variation of the $m^{loop}_{\nu}$ with $A^\nu$ in the normal
hierarchical scenario. All mass parameters are in GeV.}
\label{aneu_hie}
\end{figure}

Finally a few comments on the dependence of the one-loop contribution to the neutrino mass 
on the soft SUSY breaking parameters $Am$ and $A_\nu$. The loop contribution is always suppressed 
unless the parameter $A_\nu$ is sufficiently large as can be seen from 
Fig.\ref{aneu_hie}. As for illustration, the Yukawa couplings are chosen as 
$f^\nu_{11}=1.75\times 10^{-7}$ and $f^\nu_{22}=f^\nu_{33}=3.95\times 10^{-7}$. 
Similarly, we choose  $M_1$ = 60 GeV and $M_2$ = 120 GeV, where $M_1$ and $M_2$ are 
the $U(1)$ and $SU(2)$ gaugino mass parameters, respectively. For larger values of 
electroweak gaugino masses, the one-loop contribution would be reduced further. 
For $Am$ = -1 TeV, higher $A_\nu$ values $\sim$ 13 TeV can satisfy the current 
neutrino data. However, even if $A_\nu$ is $\sim$ 13 TeV, the quantity $A_\nu f^\nu$ is very small,
i.e., $\sim 10^{-2}$ GeV. Note that for such a choice of the parameter space, the tree level values of the 
neutrino masses are not sufficient to accommodate the three flavor global neutrino data. 
Increasing the value of $Am$ requires relatively smaller value of $A_\nu$ 
($\sim$ 7 TeV) to reproduce the neutrino data. It is very important to point out that, for a 
fixed $\lambda_s$ and $\lambda_{Ni}$ one cannot increase the trilinear coupling parameter $Am$ 
to an arbitrary high value as the right-chiral sneutrinos may turn out to be tachyonic. Thus, a 
relatively larger soft trilinear parameter $A_\nu$ is required to enhance the one-loop contribution 
to neutrino masses.
The requirement of a large $A_\nu$ can be understood from the following discussion.
\begin{itemize}
\item
%The mass splitting in the left handed sneutrinos and consequently 
The one-loop
contribution to the neutrino mass originating from the mass splitting in the
left-handed sneutrinos depends on the parameters $\mu$, $A_\nu$ and $B_N$ 
as can be seen from Eqs.(\ref{loophabermass1}) and (\ref{sneutrino-mass-splitting}).  %respectively. 
It has been argued in Ref.\cite{grossman-haber}, that in order to have 
the one-loop contribution to the neutrino mass comparable to its tree level value, the
ratio $\Delta m_{{\tilde \nu}_{ij}}/m_\nu$ should be $\sim 10^3$.
\item
Substituting $\mu = \lambda_H v_s$ and the expression for $B_N$ from Eq.(\ref{exprn-bn-mn}), 
we may write ${\Delta m_{\tilde \nu}}_{ij}/m_\nu \simeq 2(A_\nu+ \lambda_H v_s 
\cot\beta - (\frac{1}{4}\lambda_s v_s +Am/2+ \lambda_H v_d v_u/2 v_s)
(1+(A_\nu+ \lambda_H v_s \cot\beta)^2)/M^2_R)/m_{\tilde \nu}$. 
We can see from the above expression that one may increase either $Am$ or $A^\nu$ parameter 
to enhance the one-loop contribution to make it countable. But in the present context, raising 
the soft parameter $Am$ alone would not serve the purpose. This is because the VEV $v_s$ increases 
significantly with $|Am|$ (\emvide Table. \ref{tablevs}). Thus there is always a partial 
cancellation between different terms in the above expression for the left sneutrino mass splitting. 
In particular, the effective bilinear coupling $B_N$ is reduced because of this partial cancellation. 
In addition, we choose the sign of the coupling $\lambda_H$ as negative in order to determine the correct 
global minima. This also causes a partial cancellation between various terms, but to a lesser extent. 
Considering this cancellation effect in mind, it is easy to check that the ratio ${\Delta m_{\tilde
\nu}}_{ij}/m_\nu$ always reside near the value $\sim 10$ with the soft parameters $A^\nu$ and $B_N$ around 
the TeV scale.
\item
Now, as mentioned above, the trilinear coupling parameter $Am$ is restricted if one
does not want the right chiral sneutrinos to become tachyonic. Of course
this depends on the choice of the soft ``Dirac" mass term $m_{{\tilde N}{\tilde N}^*}$ 
of the $\tilde N$s, which we have chosen to have a quite moderate value (300 GeV) in this case.
However, the parameter $A^\nu$ can be pushed to a reasonably high value without affecting any
other results. This explains why a large $A^\nu$  parameter is required to make the one-loop 
contribution to the neutrino mass comparable to its tree level value.
\end{itemize}

%%%%%%%%%%%%%%%%%%%%%%%%%%%%%%%%%%%%%%%%%%%%%%%%%%%%%%%%%%%%%%%%%%%%%%%%
\section{Signatures at LHC}
It is extremely important to investigate the possible signatures of this TeV scale seesaw mechanism at the LHC.
One of the search strategies could be to produce the right-handed neutrino $N$
(or the corresponding right-handed sneutrino ${\tilde N}$) with a large enough cross-section and 
then look at the decay branching ratios in different available modes. However, in this type of models the
production of TeV scale right-handed neutrinos (or sneutrinos) at the LHC is suppressed\footnote{A very recent 
analysis along with the discovery potential at the LHC is presented in Ref.\cite{Han-Atre}.} by the light
neutrino mass \cite{neutrino_mass_LHC_review}. Nevertheless, it is possible
to construct models where the production mechanism of the right-handed neutrino (sneutrino) can be 
decoupled from the neutrino mass generation. For example, extended gauge symmetries 
such as $U(1)_{B-L}$ or $SU(2)_R$ may offer extra gauge bosons near 
the TeV scale whose couplings to quarks
and the right-handed neutrino (sneutrino) are unsuppressed \cite{mohapatra-marshak}. In such 
models a single or a pair of right-handed neutrinos can be produced with large cross sections leading to 
dilepton signals (same-sign) with no missing energy 
(see the first reference of \cite{other-early-seesaw} and
 \cite{keung-senjanovic, dittmar-etal, 
ferrari-etal, gninenko-etal}), trilepton signals \cite{del-aguila-etal} or four-lepton signals 
\cite{valle-4lepton,del-aguila-etal-2,santosh-katri-shaaban-okada-prl}.

In the context of the present model the left-sneutrino ``Majorana" mass term can lead to 
oscillation between the left-chiral sneutrino and the corresponding anti-sneutrino \cite{hirsch-klapdor-kovalenko,
grossman-haber, dedes-haber-rosiek}. This can be interpreted as the observation of a sneutrino decaying into a final-state 
with a ``wrong-sign" charged lepton. In order to have a large oscillation probability the total decay width $\Gamma$ 
of the sneutrino/antisneutrino and the mass splitting $\Delta m$ must be of the same order. Since $\Delta m$ is constrained
by the neutrino data, one needs a very small total decay width of the sneutrino/antisneutrino. It has been shown in 
\cite{grossman-haber} that this can be achieved in a scenario where the lighter stau is long-lived and the left-chiral sneutrino
can only have 3-body decay modes involving the lighter stau in the final states. This can lead to signals such as like-sign dileptons,
single charged lepton plus like-sign di-staus (leading to heavily ionizing charged tracks) or like-sign di-stau charged tracks
at future linear colliders \cite{grossman-haber,choietal,honkavaara-huitu-roy} or at the LHC \cite{honkavaara1}. The resulting 
charge asymmetry of the final states can be measured to get an estimate of the sneutrino-antisneutrino oscillation probability 
\cite{honkavaara1}. In addition, for a very small sneutrino decay width one can also observe a displaced vertex in the detector. 
However, a detailed study of such signals in the context of the present model is beyond the scope of the present paper. 

In comparison, now we discuss briefly the signatures of R-parity violating models in general.
In models with spontaneous violation of R-parity, the singlet sneutrino
vacuum expectation value leads to the existence of a Majoron which is an additional source of missing energy. This can change the
decay pattern of the lightest Higgs and the lightest neutralino with the corresponding signatures at the LHC. For more details
and the relevant references the reader is referred to Ref.\cite{neutrino_mass_LHC_review}. In the case of bilinear R-parity 
violation, the ratios of certain decay branching ratios of the LSP show very nice correlation with the neutrino mixing angles. 
This can lead to very interesting signatures at the LHC where comparable numbers of events with muons and taus, respectively, can
be observed in the final state \cite{Roy-Mukhopadhyaya-Vissani, choi-chun-kang-lee, romao-diaz-hirsch-porod-valle, 
datta-mukhopadhyaya-vissani, porod-hirsch-romao-valle, chun-jung-kang-park, jung-kang-park-chun}.  

From the above discussion we see that the canonical type-I supersymmetric seesaw case that we have considered in this paper has 
characteristic signatures which can be tested at the LHC. At the same time one can also distinguish the predictions of this model 
with those of the models with spontaneous or bilinear R-parity violating 
scenarios.  
%%%%%%%%%%%%%%%%%%%%%%%%%%%%%%%%%%%%%%%%%%%%%%%%%%%%%%%%%%%%%%%%%%%%%%%%

\section{Conclusions}
We have studied the neutrino masses and mixing in an R-parity conserving supersymmetric 
standard model with three right handed neutrino superfields ${\hat N}_i$ and another gauge 
singlet superfield ${\hat S}$. This model is similar to the next-to-minimal supersymmetric standard model
(NMSSM), where the scalar component of ${\hat S}$ gets a VEV to generate a $\mu$-term of 
correct order of magnitude. In addition, the same VEV also generates TeV scale Majorana masses
for the right handed neutrinos. The small neutrino masses are generated at the tree level by
the usual seesaw mechanism at the TeV scale. We also calculate the one-loop contribution to the
neutrino mass matrix and investigate the constraints on the model parameters to produce the
tri-bimaximal pattern of neutrino mixing for three different neutrino 
mass hierarchies. Neutrino 
mass matrix gets contribution at the one-loop level controlled by the sneutrino ``Majorana" mass 
terms. We show that the one-loop contribution can be important for certain choices
of the soft SUSY breaking parameters. This we have demonstrated by evaluating the one-loop 
contribution in two different ways. In particular, we observe that the one-loop contributions can 
be significant when the soft SUSY breaking trilinear parameter $A_\nu f^\nu$ is 
$\sim {\cal O} (10^{-3}$ GeV) with $A_\nu \sim$ 10 TeV. This observation is quite robust and does not 
change much if one introduces a small $\theta_{13}$ in the neutrino sector. Our choice of neutrino
Yukawa couplings also predict vanishing contributions to the lepton flavor violating processes
$\mu \rightarrow e \gamma$ and $\tau \rightarrow e \gamma$ as well as an extremely suppressed 
contribution to $\tau \rightarrow \mu \gamma$.

As has been stated earlier, it is also possible to have non-zero vacuum expectation 
values for the left and right chiral sneutrinos. In that case, R-parity is 
violated spontaneously. The neutrino mass matrix can have contributions from two different sources,
namely, the effective bilinear R-parity violating interactions and the TeV-scale seesaw mechanism. 
One-loop contributions to the neutrino mass matrix can be very important in this case too. However,
the tree level and one-loop calculations are rather involved and require a separate discussion 
altogether. We plan to present these results in a subsequent paper\cite{debottam-future}. 

The characteristic signatures of this model at the LHC include like-sign dilepton (without missing energy),
trilepton or four lepton final states as well as single lepton plus two heavily ionizing charged tracks or
only two heavily ionizing charged tracks stemming from long-lived staus. By looking at these signals one
can possibly distinguish this model from the models of spontaneous or bilinear R-parity violation.  

\noindent
{\bf Acknowledgments}\\ 
DD thanks P2I, CNRS for the support received as a post-doctoral fellow. We thank  Pradipta Ghosh, 
Biswarup Mukhopadhyaya and Subhendu Rakshit for fruitful discussions. DD would also like to thank 
Asmaa Abada and Gregory Moreau for some valuable comments and suggestions. He is also thankful to the 
Department of Theoretical Physics, Indian Association for the Cultivation of Science 
(IACS), where a part of this work has been done.


\begin{thebibliography}{1} 

\bibitem{r-parity-early-references}
For early references, see, P.~Fayet, \NPB {\bf 90}, 104 (1975); \PLB {\bf 69}, 489 (1977);
   G.~R.~Farrar and P.~Fayet, \PLB {\bf 76}, 575 (1978);
  C.~S.~Aulakh and R.~N.~Mohapatra, \PLB {\bf 119}, 136 (1982);
    L.~J.~Hall and M.~Suzuki, \NPB {\bf 231}, 419 (1984);
    I.~H.~Lee, \PLB {\bf 138}, 121 (1984);
   I.~H.~Lee, \NPB {\bf 246}, 120 (1984);
  G.~G.~Ross and J.~W.~F.~Valle, \PLB {\bf 151}, 375 (1985);
     J.~R.~Ellis, G.~Gelmini, C.~Jarlskog, G.~G.~Ross and J.~W.~F.~Valle,
   \PLB {\bf 150}, 142 (1985);
  A. Masiero and J.W.F. Valle, \PLB {\bf 251}, 273 (1990).

\bibitem{r-parity-review}
For reviews on R-parity violation, see, e.g., R. Barbier {\it et al.}, Phys.
Rep. {\bf 420}, 1 (2005); M. Chemtob, Prog. Part. Nucl. Phys. {\bf 54},
71 (2005).

\bibitem{romao-santos-valle}
J.C. Romao,  C.A. Santos, J.W.F. Valle,
% {\it {How to spontaneously break R-parity}},
  \PLB{\bf 288}, 311 (1992).

\bibitem{giudice-masiero-pietroni-riotto}
G.F. Giudice, A. Masiero, M. Pietroni, A. Riotto,
% {\it {The Supersymmetric singlet majoron}},
  \NPB{\bf 396}, 243 (1993) [arXiv:hep-ph/9209296].

\bibitem{umemura-yamamoto}
I. Umemura and K. Yamamoto,
% {\it {Neutrinos in the supersymmetric singlet majoron model}},
  \NPB{\bf 423}, 405 (1994).

%%%%%%%%%%%%%%%%%%%%%%%%%%%%%%%%%%%%%%%%%%%%%%%%%%%%%%%%%%%%%%%%%%%%%%%%%%
%%%%%%%%%%%%%%%%%              RPV SIGNALS     %%%%%%%%%%%%%%%%%%%%%%%%%%%
%%%%%%%%%%%%%%%%%%%%%%%%%%%%%%%%%%%%%%%%%%%%%%%%%%%%%%%%%%%%%%%%%%%%%%%%%%
\bibitem{gonzalez-garcia-romao-valle}
M.C. Gonzalez-Garcia, J.C. Romao, J.W.F. Valle,
 %{\it {Spontaneous R-parity breaking at hadron supercolliders}},
  \NPB{\bf 391}, 100 (1993).

\bibitem{adhikari-mukhopadhyaya}
R. Adhikari and B. Mukhopadhyaya,
 %{\it {Distinctive signals of spontaneous R-parity breaking at LEP-2}},
  \PLB{\bf 378}, 342 (1996) [arXiv:hep-ph/9601382]; \\
   {\it Erratum-ibid.}  {\bf B384}, 492 (1996).

\bibitem{hirsch-vicente-porod}
M. Hirsch, A. Vicente, W. Porod,
 %{\it {Spontaneous R-parity violation: Lightest neutralino decays and neutrino mixing angles at future colliders}},
  \PRD{\bf 77}, 075005 (2008) [arXiv:0802.2896].

\bibitem{Roy-Mukhopadhyaya-Vissani}
  B.~Mukhopadhyaya, S.~Roy and F.~Vissani,
 % {\it{Correlation between neutrino oscillations and collider signals of
 % supersymmetry in an R-parity violating model}},
  \PLB{\bf 443}, 191 (1998) [hep-ph/9808265].

\bibitem{choi-chun-kang-lee}
S.Y. Choi, E. J. Chun, S. K. Kang, J. S. Lee,
 %{\it {Neutrino oscillations and R-parity violating collider signals}},
  \PRD{\bf 60}, 075002 (1999) [hep-ph/9903465].

\bibitem{romao-diaz-hirsch-porod-valle}
J.C. Romao, M.A. Diaz, M. Hirsch, W. Porod, J.W.F Valle,
 %{\it {A Supersymmetric solution to the solar and atmospheric neutrino problems}},
  \PRD{\bf 61}, 071703 (2000) (Rapid Communications) [hep-ph/9907499].

\bibitem{datta-mukhopadhyaya-vissani}
A. Datta, B. Mukhopadhyaya and F. Vissani,
 %{\it {Tevatron signatures of an R-parity violating supersymmetric theory}},
  \PLB{\bf 492}, 324 (2000) [hep-ph/9910296].

\bibitem{porod-hirsch-romao-valle}
W. Porod, M. Hirsch, J. Romao and J.W.F. Valle,
 %{\it {Testing neutrino mixing at future collider experiments}},
  \PRD{\bf 63}, 115004 (2001) [hep-ph/0011248].

\bibitem{chun-jung-kang-park}
E. J. Chun, D-W. Jung, S. K. Kang, J. D. Park,
 %{\it {Collider signatures of neutrino masses and mixing from R parity violation}},
  \PRD{\bf 66}, 073003 (2002) [hep-ph/0206030].

\bibitem{jung-kang-park-chun}
D-W. Jung, S. K. Kang, J. D. Park, E. J. Chun,
 %{\it {Neutrino oscillations and collider test of the R-parity violating minimal supergravity model}},
  \JHEP{\bf 08}, 017 (2004)  [arXiv:hep-ph/0407106].

%%%%%%%%%%%%%%%%%%%%%%%%%%%%%%%%%%%%%%%%%%%%%%%%%%%%%%%%%%%%%%%%%%%%%%%%%%
\bibitem{original-seesaw}
  P.~Minkowski,
  Phys. Lett. B {\bf 67}, 421 (1977);
M.~Gell-Mann, P.~Ramond and R.~Slansky,
Published in Supergravity, P. van Nieuwenhuizen $\&$ D.Z. Freedman (eds.),
North Holland Publ. Co., 1979. Published in Stony Brook Wkshp.1979:0315 (QC178:S8:1979);
T.~Yanagida,
{\it{in \emph{Proceedings of the Workshop on the Unified Theory
and the Baryon Number in the Universe} (O.~Sawada and A.~Sugamoto, eds.), KEK,
  Tsukuba, Japan, 1979, p.~95}};
S.~L. Glashow,
  \emph{Proceedings of the 1979 Carg{\`e}se Summer Institute on Quarks and
  Leptons} (M.~L{\'e}vy, J.-L. Basdevant, D.~Speiser, J.~Weyers,
R.~Gastmans, and M.~Jacob, eds.), Plenum Press, New York, 1980, pp.~687--713;
  R.~N.~Mohapatra and G.~Senjanovic,
  Phys. Rev. Lett. {\bf 44}, 912 (1980).

\bibitem{other-early-seesaw}
J. Schechter and J.W.F. Valle, \PRD {\bf 22}, 2227 (1980);\PRD {\bf 25}, 774 (1982);
T.P. Cheng and L.-F. Li, \PRD {\bf 22}, 2860 (1980).

\bibitem{grossman-haber}
  Y.~Grossman and H.~E.~Haber,
  Phys.\ Rev.\ Lett.\  {\bf 78}, 3438 (1997)
[arXiv:hep-ph/9702421].
\bibitem{susy-seesaw}
S.~F.~King,
  Phys.\ Lett.\  B {\bf 439}, 350 (1998)
  [arXiv:hep-ph/9806440].
%  S.~Davidson and S.~F.~King,
%  Phys.\ Lett.\  B {\bf 445}, 191 (1998)
%  [arXiv:hep-ph/9808296].
\bibitem{davidson-king}
  S.~Davidson and S.~F.~King,
  %neutrino,''
  Phys.\ Lett.\  B {\bf 445}, 191 (1998)
  [arXiv:hep-ph/9808296].

\bibitem{NMSSM_review1}
 M.~Maniatis,
  arXiv:0906.0777 [hep-ph];
U.~Ellwanger, C.~Hugonie and A.~M.~Teixeira,
  arXiv:0910.1785 [hep-ph].

\bibitem{NMSSM_review2}
P.N.~Pandita, \Journal{\PLB}{318}{338}{1993};
P.N.~Pandita, \Journal{\ZPC}{59}{575}{1993};
U.~Ellwanger {\it et al.}, {\it Z. Phys.} C {\bf 67}, 665 (1995);
U.~Ellwanger {\it et al.}, \Journal{\NPB}{492}{21}{1997};
U.~Ellwanger and C.~Hugonie, {\it Eur. Phys. J.} C {\bf 13}, 681 (2000);
A.~Dedes {\it et al.}, \Journal{\PRD}{63}{055009}{2001};
U.~Ellwanger {\it et al.}, {\tt arXiv:hep-ph/0111179};
D.~G.~Cerdeno {\it et al.}, \Journal{\JHEP}{0412}{048}{2004};
U.~Ellwanger {\it et al.}, \Journal{\JHEP}{0507}{041}{2005};
U.~Ellwanger and C.~Hugonie, \Journal {\PLB}{623}{93}{2005};
G.~B\'elanger {\it et al.}, \Journal{JCAP}{0509}{001}{2005};
%J.~F.~Gunion, D.~Hooper and B.~McElrath, \Journal{\PRD}{73}{ 015011}{2006};
 A.~Djouadi {\it et al.},
  %``Benchmark scenarios for the NMSSM,''
  JHEP {\bf 0807}, 002 (2008);
%  [arXiv:0801.4321 [hep-ph]];
%F.~Domingo and U.~Ellwanger,
%  JHEP {\bf 0807}, 079 (2008);
%  [arXiv:0806.0733 [hep-ph]];
  %%CITATION = JHEPA,0807,079;
A.~Djouadi, U.~Ellwanger and A.~M.~Teixeira,
  JHEP {\bf 0904}, 031 (2009).
%  [arXiv:0811.2699 [hep-ph]].

\bibitem{chemtob-pandita}
P.N. Pandita and P.F. Paulraj, Phys. Lett. {\bf B462}, 294 (1999);
P.N. Pandita, Phys. Rev. D{\bf 64}, 056002 (2001);
M. Chemtob and P.N. Pandita, Phys. Rev. D {\bf 73}, 055012 (2006);
A. Abada and G. Moreau, J. High Energy Phys., {\bf 08}, 044 (2006).

\bibitem{abada-bhattacharyya-moreau}
A. Abada, G. Bhattacharyya, and G. Moreau, Phys. Lett. {\bf B642}, 503 (2006).

\bibitem{Kitano-2001}
  R.~Kitano and K.~y.~Oda,
  Phys.\ Rev.\  D {\bf 61}, 113001 (2000)
  [arXiv:hep-ph/9911327].

\bibitem{ellis89} P. Fayet, Nucl. Phys. {\bf B90}, 104 (1975); R. Barbieri, S. Ferrara, and 
C.A. Savoy, Phys. Lett. {\bf 119B}, 343 (1982); J. Ellis, J.F. Gunion, H.E.  Haber, L. Roszkowski,
and F. Zwirner, Phys. Rev. D{\bf 39}, 844 (1989).

\bibitem{Strumia}
E.~K.~Akhmedov,
  %``Neutrino physics,''
arXiv:hep-ph/0001264;
A.~Strumia and F.~Vissani, Nucl. Phys. {\bf B726}, 294 (2005);
%\bibitem{Akhmedov:1999uz}
  %%CITATION = HEP-PH/0001264;%%
R.~N.~Mohapatra and A.~Y.~Smirnov,
  %``Neutrino Mass and New Physics,''
hep-ph/0603118.

\bibitem{neutrino_sum}
  E.~Komatsu {\it et al.}  [WMAP Collaboration],
{\it {Five-Year Wilkinson Microwave Anisotropy Probe (WMAP)
  Observations:Cosmological Interpretation}},
  Astrophys. J. Suppl. {\bf 180}, 330 (2009)
  [arXiv:0803.0547 [astro-ph]].

\bibitem{neutrino_mass}
H.~V.~Klapdor-Kleingrothaus, I.~V.~Krivosheina, A.~Dietz and O.~Chkvorets,
Phys.\ Lett.\  B {\bf 586}, 198 (2004)
[arXiv:hep-ph/0404088];
H.~V.~Klapdor-Kleingrothaus and I.~V.~Krivosheina,
Mod. Phys. Lett.  A {\bf 21}, 1547 (2006); C.~Arnaboldi {\it et al.}  [CUORICINO Collaboration],
  Phys. Rev. C {\bf 78}, 035502 (2008)
  [arXiv:0802.3439 [hep-ex]].

\comment{\bibitem{neu-recent-data}
  T.~Schwetz,
  %``Global fits to neutrino oscillation data,''
  Phys.\ Scripta {\bf T127}, 1 (2006)
  [arXiv:hep-ph/0606060].

\bibitem{PDG}
 W.~M.~Yao {\it et al.}  [Particle Data Group],
  %``Review of particle physics,''
  J.\ Phys.\ G {\bf 33}, 1 (2006).
  %%CITATION = JPHGB,G33,1;%%
\bibitem{data}
 M.~C.~Gonzalez-Garcia and M.~Maltoni,
  %``Phenomenology with Massive Neutrinos,''
  arXiv:0704.1800 [hep-ph].}
\bibitem{new-neu-recent-data}
T.~Schwetz, M.~A.~Tortola and J.~W.~F.~Valle,
  %``Three-flavour neutrino oscillation update,''
  New J.\ Phys.\  {\bf 10}, 113011 (2008)
  [arXiv:0808.2016 [hep-ph]].

\bibitem{hps}
P.F. Harrison, D.H. Perkins and W.G. Scott, Phys. Lett. B {\bf 530},
167 (2002).

\bibitem{cerdenoall}
D.~G.~Cerdeno, C.~Munoz and O.~Seto,
  %``Right-handed sneutrino as thermal dark matter,''
  Phys.\ Rev.\  D {\bf 79}, 023510 (2009)
  [arXiv:0807.3029 [hep-ph]]; D.~G.~Cerdeno and O.~Seto,
  %``Right-handed sneutrino dark matter in the NMSSM,''
  JCAP {\bf 0908}, 032 (2009)
  [arXiv:0903.4677 [hep-ph]].



\bibitem{katri-mariana-timo}
M. Frank, K. Huitu, and T. R\"{u}ppell, Eur. Phys. J. {\bf C52}, 413 (2007).

\bibitem{manimala}
M.~Mitra,
  %``Spontaneous R-Parity Violation, $A_4$ Flavor Symmetry and Tribimaximal
  %Mixing,''
  arXiv:0912.5291 [hep-ph].

\bibitem{moreau-dark}
C.~C.~Jean-Louis and G.~Moreau,
  %``Dark matter and neutrino masses in the R-parity violating NMSSM,''
  arXiv:0911.3640 [hep-ph].

\bibitem{munoz}
D. E. L\'{o}pez-Fogliani and C. Mu\~{n}oz, Phys. Rev. Lett {\bf 97}, 041801 (2006);
N. Escudero, D.E. L\'{o}pez-Fogliani, C. Mu\~{n}oz, and R.R. de Austri, J. High Energy
Phys. {\bf 12}, 099 (2008); J. Fidalgo, D.E. L\'{o}pez-Fogliani, C. Mu\~{n}oz, and
R. Ruiz de Austri, J. High Energy Phys. {\bf 08}, 105 (2009) [arXiv:0904.3112].

\bibitem{ghosh-roy}
P. Ghosh and S. Roy, J. High Energy Phys. {\bf 04}, 069 (2009).

\bibitem{ghosh-roy-loop}
P.~Ghosh, P.~Dey, B.~Mukhopadhyaya and S.~Roy,
  %``Radiative contribution to neutrino masses and mixing in $\mu\nu$SSM,''
J. High Energy Phys. {\bf 05}, 087 (2010)
  arXiv:1002.2705 [hep-ph].

\bibitem{hirsch-vicente}
A. Bartl, M. Hirsch, S. Liebler, W. Porod, and A. Vicente, J. High Energy Phys. 
{\bf 05}, 120 (2009) [arXiv:0903.3596].

\bibitem{Mukhopadhyaya:2006is}
  B.~Mukhopadhyaya and R.~Srikanth,
  %``Bilarge neutrino mixing in R-parity violating supersymmetry: The role  of
  %right-chiral neutrino superfields,''
  Phys.\ Rev.\  D {\bf 74}, 075001 (2006)
  [arXiv:hep-ph/0605109].

\bibitem{higher-dimension}
N.~Arkani-Hamed, L.~J.~Hall, H.~Murayama, D.~Tucker-Smith and N.~Weiner,
  %``Small neutrino masses from supersymmetry breaking,''
  Phys.\ Rev.\  D {\bf 64}, 115011 (2001)
  [arXiv:hep-ph/0006312];
N.~Arkani-Hamed, L.~J.~Hall, H.~Murayama, D.~Tucker-Smith and N.~Weiner,
  %``Neutrino masses at v**(3/2),''
  arXiv:hep-ph/0007001; F.~Borzumati and Y.~Nomura,
  %``Low-scale see-saw mechanisms for light neutrinos,''
  Phys.\ Rev.\  D {\bf 64}, 053005 (2001)
  [arXiv:hep-ph/0007018]; J.~March-Russell and S.~M.~West,
  %``A simple model of neutrino masses from supersymmetry breaking,''
  Phys.\ Lett.\  B {\bf 593}, 181 (2004)
  [arXiv:hep-ph/0403067]; B.~Mukhopadhyaya, P.~Roy and R.~Srikanth,
  %``Bilarge neutrino mixing from supersymmetry with high scale
  %nonrenormalizable interactions,''
  Phys.\ Rev.\  D {\bf 73}, 035003 (2006);
J. March-Russell, C. McCabe, M. McCullough,
 %Neutrino-Flavoured Sneutrino Dark Matter.
 \JHEP {\bf 1003}, 108 (2010) arXiv:0911.4489 [hep-ph].

\bibitem{debottam-future}
D. Das and S. Roy, in preparation.

\bibitem{tev-seesaw}
Y.~Grossman and H.~E.~Haber,
  %``The would-be majoron in R-parity violating supersymmetry,''
  Phys.\ Rev.\  D {\bf 67}, 036002 (2003)
  [arXiv:hep-ph/0210273].


  %%CITATION = PHRVA,D61,113001;%%

\bibitem{hierarchy}
E. Ma, Phys. Rev. D {\bf 66}, 117301 (2002); A. Joshipura, Proc. Indian
Natl. Sci. Acad {\bf A 70}, 223 (2004).

\bibitem{Han-Atre}
  A.~Atre, T.~Han, S.~Pascoli and B.~Zhang,
% {\it{The Search for Heavy Majorana Neutrinos}},
   J. High Energy Phys. {\bf 05}, 030 (2009) arXiv:0901.3589 [hep-ph].

\bibitem{neutrino_mass_LHC_review}
P. Nath {\it et al.}, Nucl. Phys. Proc. Suppl. {\bf 200--202}, 185--417 (2010) 
arXiv:1001.2693 [hep-ph].

\bibitem{mohapatra-marshak}
R.N. Mohapatra and R.E. Marshak, Phys. Rev. Lett. {\bf 44}, 1316 (1980).

%\bibitem{schechter-valle}
%J. Schechter and J.W.F. Valle, Phys. Rev. D{\bf 22}, 2227 (1980) in Ref. \cite{other-early-seesaw}.

\bibitem{keung-senjanovic}
W.-Y. Keung and G. Senjanovic, Phys. Rev. Lett. {\bf 50}, 1427 (1983). 

\bibitem{dittmar-etal}
M. Dittmar {\it et al.}, \NPB {\bf 332}, 1 (1990).

\bibitem{ferrari-etal}
A. Ferrari {\it et al.}, Phys. Rev. D{\bf 62}, 013001 (2000).

\bibitem{gninenko-etal}
S.N. Gninenko, M.M. Kirsanov, N.V. Krasnikov, and V.A. Matveev, Phys. Atom. Nucl. {\bf 70}, 441 (2007).

\bibitem{del-aguila-etal}
F. del Aguila, J.A. Aguilar-Saavedra, and J. de Blas, arXiv:0910.2720.

\bibitem{valle-4lepton}
J.W.F. Valle, Phys. Lett. {\bf B196}, 157 (1987).

\bibitem{del-aguila-etal-2}
F. del Aguila and J.A. Aguilar-Saavedra, J. High Energy Phys. {\bf 11}, 072 (2007).

\bibitem{santosh-katri-shaaban-okada-prl}
For a more recent reference, see e.g., K. Huitu, S. Khalil, H. Okada, and S.K. Rai, Phys. Rev. Lett. {\bf 101}, 
181802 (2008).

\bibitem{hirsch-klapdor-kovalenko}
M. Hirsch, H.V. Klapdor-Kleingrothaus, and S.G. Kovalenko, Phys. Lett. {\bf B398}, 311 (1997).

\bibitem{dedes-haber-rosiek}
A. Dedes, H.E. Haber, and J. Rosiek, J. High Energy Phys. {\bf 11}, 059 (2007).

\bibitem{choietal}
 K. Choi, K. Hwang, and W.Y. Song, Phys. Rev. Lett. {\bf 88}, 141801 (2002).

\bibitem{honkavaara-huitu-roy}
T. Honkavaara, K. Huitu, and S. Roy, Phys. Rev. D{\bf 73}, 055011 (2006).

\bibitem{honkavaara1}
D.K. Ghosh, T. Honkavaara, K. Huitu, S. Roy, \PRD {\bf 79}, 055005 (2009); arXiv:1005.1802.


\end{thebibliography}
\end{document}